%% file: Manuscript_Final.tex
\documentclass[preprint,12pt,3p]{elsarticle}

\input{preamble.tex}

\biboptions{sort&compress}

\begin{document}

\textbf{Highlights}
\begin{itemize}
  \item B\'ezier extraction based IGA is successfully implemented for geometrically nonlinear static and dynamic analyses of smart FG plate structure.
  \item Two porosity distributions and three GPL dispersion patterns along the thickness direction is examined.
 \item Influence of porosity coefficient, weight fraction of GPLs and external electrical voltage are investigated.
  \item Active control of nonlinear dynamic responses via sensor and actuator layers is shown.
  \item The best reinforcement performance of piezoelectric FG porous plate structures is explored.
\end{itemize}

\begin{frontmatter}

  \clearpage
  \title{Analysis and active control of geometrically nonlinear responses of smart FG porous plates with graphene nanoplatelets reinforcement based on B\'ezier extraction of NURBS}
  \date{} 
  \author[label1]{Nam V. Nguyen}
  \author[label2]{Lieu B. Nguyen}
  \author[label3]{H. Nguyen-Xuan \corref{cor}}
  \ead{ngx.hung@hutech.edu.vn}
   \author[label1]{Jaehong Lee\corref{cor}}
  \ead{jhlee@sejong.ac.kr}
  \address[label1]{Department of Architectural Engineering, Sejong University, 209 Neungdong-ro, Gwangjin-gu, Seoul, 05006, Republic of Korea}
  \address[label2]{Faculty of Civil Engineering, University of Technology and Education Ho Chi Minh City, Vietnam}
  \address[label3]{CIRTech Institute, Ho Chi Minh City University of Technology (HUTECH), Ho Chi Minh City, Vietnam}
  \cortext[cor]{Corresponding author}

  \begin{abstract}
    In this paper, we propose an effective computational approach to analyze and active control of geometrically nonlinear responses of functionally graded (FG) porous plates with graphene nanoplatelets (GPLs) reinforcement integrated with piezoelectric layers. The key concept behind this work is to utilize isogeometric analysis (IGA) based on B\'ezier extraction technique and $C^0$-type higher-order shear deformation theory ($C^0$-HSDT). By applying B\'ezier extraction, the original Non-Uniform Rational B-Spline (NURBS) control meshes can be transformed into B\'ezier elements which allow us to inherit the standard numerical procedure like the standard finite element method (FEM). In this scenario, the approximation of mechanical displacement field is calculated via $C^0$-HSDT whilst the electric potential field is considered as a linear function across the thickness of each piezoelectric sublayer. The FG plate includes internal pores and GPLs dispersed into metal matrix either uniformly or non-uniformly along plate's thickness. To control responses of structures, the top and bottom surfaces of FG plate are firmly bonded with piezoelectric layers which are considered as sensor and actuator layers. The geometrically nonlinear equations are solved by Newton-Raphson iterative procedure and Newmark's integration. The influence of porosity coefficient, weight fraction of GPLs as well as external electrical voltage on geometrically nonlinear behaviors of plate structures with various distributions of porosity and GPLs are thoroughly investigated. A constant displacement and velocity feedback control approaches are then adopted to actively control geometrically nonlinear static and dynamic responses, where structural damping effect is taken into account, based on a closed-loop control with sensor and actuator layers.

  \end{abstract}

  \begin{keyword}
    Porous plate \sep Graphene nanoplatelets \sep Piezoelectric material \sep B\'ezier extraction \sep Nonlinear dynamic \sep Active control
  \end{keyword}

\end{frontmatter}
\section{Introduction}
\label{sec:introduction}

A large demand for high structural performance and multifunctionality having superior mechanical properties and chemical stability in engineering applications has been significantly increased in the last few years. With cellular structures, porous materials having outstanding properties such as excellent energy absorption, lightweight, heat resistance has been broadly employed in various engineering fields \cite{lefebvre2008porous, smith2012steel}. However, the existence of internal pores in metal matrix leads to a significant reduction in terms of structural stiffness \cite{xia2013effects}. To make up for this shortcoming, reinforcement with carbonaceous nanofillers e.g. carbon nanotubes (CNTs) \cite{iijima1991helical, liew2015mechanical} and graphene nanoplatelets (GPLs) \cite{mittal2015review} into lightweight materials is considered as an excellent and practical choice to strengthen their mechanical properties. More importantly, this reinforcement also aims to preserve their potential for lightweight structures \cite{duarte2015effective}. Compared to CNTs, GPLs have exhibited great potentials to become a good reinforcement candidate \cite{rafiee2009enhanced} since GPLs possess superior mechanical properties, a lower manufacturing cost, a larger specific surface area and two-dimensional geometry. To enhance the performance of structures, functionally graded (FG) porous structures reinforced by GPLs have been proposed in the literature to obtain the desired mechanical properties by adjusting the size and density of porosities as well as distribution of internal pores and GPLs \cite{hassani2012production,he2014preparation}. In terms of numerical analysis, numerous investigations have been carried out to examine the effect of internal pores and GPLs on structural behaviors under various conditions. Based on Timoshenko's beam theory and Ritz method, the free vibration and buckling analyses of FG porous beams reinforced with GPLs are conducted by Kitipornchai et al. \cite{kitipornchai2017free} while the nonlinear free vibration and postbuckling behaviors are reported by Chen et al. \cite{chen2017nonlinear}. In addition, the nonlinear responses of arch and ring porous structures reinforced by GPLs are also examined \cite{liu2019nonlinear, li2019analytical}. Yang et al. \cite{yang2018buckling} employed the first-order shear deformation plate theory (FSDT) and Chebyshev-Ritz method to investigate the buckling and free vibration behaviors of FG porous plate structures with GPL reinforcement. Then, the nonlinear vibration and dynamic buckling of FG porous plates reinforced with GPLs resting on the elastic foundations are considered in \cite{gao2018nonlinear, li2018nonlinear}. Shahverdi et al. \cite{shahverdi2019post} presented the postbuckling analysis for perfect/imperfect honeycomb core sandwich plate with GPL reinforcement. Based on isogeometric analysis (IGA), Li et al. \cite{li2018isogeometric} performed the static, free vibration and buckling analyses of FG porous plate structures reinforced by GPLs.\\

On the other hand, piezoelectric materials have been developed and widely employed in various engineering fields to produce smart structures. One of the essential and excellent features of this material type is the capacity transformation between mechanical and electrical energy, which is known as piezoelectric effect and converse phenomenon \cite{wang1997static}. Regarding the analysis for plate structures embedded in piezoelectric layers, numerous researches have been carried out in order to predict their behaviors in the literature \cite{he2001active, liew2003modelling,selim2016active,phung2017nonlinear}. Besides, FG carbon nanotubes reinforced composite plates (FG-CNTRC) embedded in piezoelectric layers also attracted remarkable attention of researchers. Alibeigloo \cite{alibeigloo2013static, ALIbeigloo2014free} presented the static and free vibration analyses of FG-CNTRC plate as well as cylindrical panel embedded in thin piezoelectric layers. By employing FSDT, Sharma et al. \cite{sharma2016smart} reported the active vibration control of FG-CNTRC plates carrying piezoelectric sensor and actuator layers. Based on element-free IMLS-Ritz model and HSDT, Selim et al. \cite{selim2017active,selim2019active,selim2019active1} conducted a series of works regarding the active vibration control of FG plates reinforced by CNTs or GPLs embedded in piezoelectric layers. The dynamic response of laminated CNTRC plates incorporated with piezoelectric layers is addressed by Nguyen-Quang et al. \cite{nguyen2018isogeometric} using IGA based on HSDT. In addition, Malekzade et al. \cite{malekzadeh2018vibration} examined the free vibration of FG eccentric annular plates with GPLs reinforcement and embedded in piezoelectric layers. Most recently, the FG porous plate structures reinforced by GPLs and integrated piezoelectric layers are considered by Nguyen et al. \cite{nguyen2019isogeometric} for static and dynamic analyses and Nguyen et al. \cite{nguyen2019active} for active vibration control problems. \\

It is known that the different basic functions are applied for approximation of geometry models and solutions in the framework of traditional FEM which can lead to errors in computational process. In order to tackle this issue, IGA which employed non-uniform rational B-splines (NURBS) basis as shape functions are developed by Hughes et al. \cite{hughes2005isogeometric}. The main idea of IGA is fulfilled by using the same shape functions to describe the geometry model as well as to approximate the solution field. Till now, the IGA has been successfully applied to various engineering fields \cite{nguyen2015isogeometric}. In comparison with standard FEM, the NURBS based IGA yields better accuracy and efficiency for many engineering issues, particularly for ones with complicated geometry models. The basic and review of IGA are presented in the established literature \cite{cottrell2009isogeometric, nguyen2015isogeometric}. Nevertheless, it should be emphasized that implementing NURBS based IGA approach seems not to easy as its basis functions are not confined to a unique element but instead span over the whole parametric domain. In addition, the calculation of stiffness matrix as well as force vector requires a mapping process between parametric and physical spaces. To overcome these obstacles, Borden et al. \cite{borden2011isogeometric} first proposed B\'ezier extraction technique for NURBS by using Bernstein polynomials basis which is similar to the Lagrangian shape functions. Accordingly, B\'ezier extraction technique decomposes NURBS basis function into a linear combination of Bernstein polynomials basis and obtains $C^0$ continuity of isogeometric B\'ezier elements. By employing Bernstein polynomials like basis functions in B\'ezier extraction concept, the implementation of IGA becomes compatible with the traditional FEM while higher degree shape functions are efficiently exploited. As a result, the IGA approach can be easily embedded in most existing FEM codes while its advantages are still kept naturally. Several research works based on B\'ezier extraction concept can be found in the literature \cite{scott2011isogeometric, irzal2014isogeometric, lai20173}. \\

From the above literature review, it can be observed that the mentioned researches have only focused on plate structures integrated with piezoelectric layers in which core layers are often made of FGM or FG-CNTRC. Furthermore, the geometrically nonlinear analyses of piezoelectric FG plates under various dynamic loads are still somewhat limited. In the view of practical applications of high performance and smart engineering structures, the geometrically nonlinear static and dynamic responses of piezoelectric FG plate structures should be thoroughly examined to support the structural design process. In addition, the exploiting and integrating the outstanding features of IGA into the available FEM framework should be more considered in the literature because of the excellent computational effectiveness of them. This work aims to fill the research gap in the literature regarding the analysis and control of geometrically nonlinear responses of piezoelectric FG plates having the core layer composed of porous materials reinforced by GPLs. Accordingly, B\'ezier extraction technique based IGA in conjunction with $C^0$-HSDT is fully exploited in this study. By using the B\'ezier extraction, the IGA retains the element structure which allows the IGA approach to integrate conveniently into the existing FEM routine. The active control for geometrically nonlinear static and dynamic responses of FG plates with structural damping effect using a closed loop control via sensor and actuator layers is presented. In this study, the core layers are instituted by a combination of two porosity distributions and three GPL dispersion patterns across plate's thickness, respectively while the piezoelectric layers are perfectly bonded on both top and bottom surfaces of core layer. The Newmark's integration scheme in conjunction with Newton-Raphson technique is utilized for nonlinear analyses. Several numerical verifications are first conducted in order to demonstrate the accuracy and efficiency of the proposed approach. Numerically obtained results show that the responses of piezoelectric FG are significantly influenced by porosity coefficient, weight fraction of GPLs, distribution of porosity and GPLs into metal matrix as well as input voltages. More importantly, the best reinforcement performance for piezoelectric FG porous plate structures reinforced by GPLs is discovered in this study. \\

The outline of this work as follows. Section \ref{sec:Theoretical_formulations} provides theoretical formulations including material models, $C^0$-type HSDT, constitutive equations and weak form. The brief on B\'ezier extraction based IGA and discrete system equations are given in Section \ref{sec:IGA_Section}. The active control algorithm is described in Section \ref{sec:Active_Control}. Section \ref{sec:num_exam} presents numerical investigations for the geometrically nonlinear static and dynamic analyses along with active control. The paper is closed with some concluding remarks which can be found in Section \ref{sec:Conclusions}.\\

\section{Theoretical formulations}
\label{sec:Theoretical_formulations}
\subsection{Material models}

\begin{figure}[h!]
  \centering
  \includegraphics[trim=5cm 9.2cm 7cm 2.5cm,clip=true,scale=0.7]{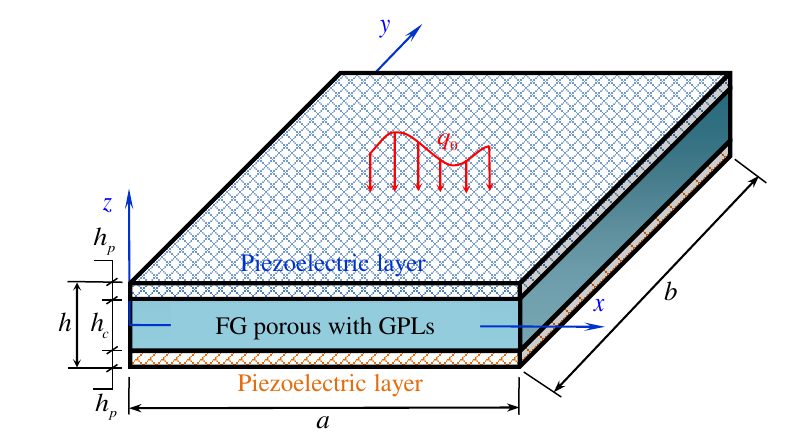}
  \caption{Model of a sandwich FG plate structure under a load $q_0$. The core layer and two face layers are the porous materials with GPL reinforcement and piezoelectric materials, respectively. The plate has the length $a$, the width $b$ and the total thickness $h=h_c+2h_p$, where $h_c$ and $h_p$ are the thickness of core and each face layer, respectively.}
  \label{fig:Model_plates}
\end{figure}

\begin{figure}[h!]
  \centering
  \begin{subfigure}[b]{1.0\textwidth}
    \begin{subfigure}[b]{0.495\textwidth}
      \centering
      \includegraphics[trim=3cm 2.5cm 7cm 2.5cm,clip=true,scale=0.3]{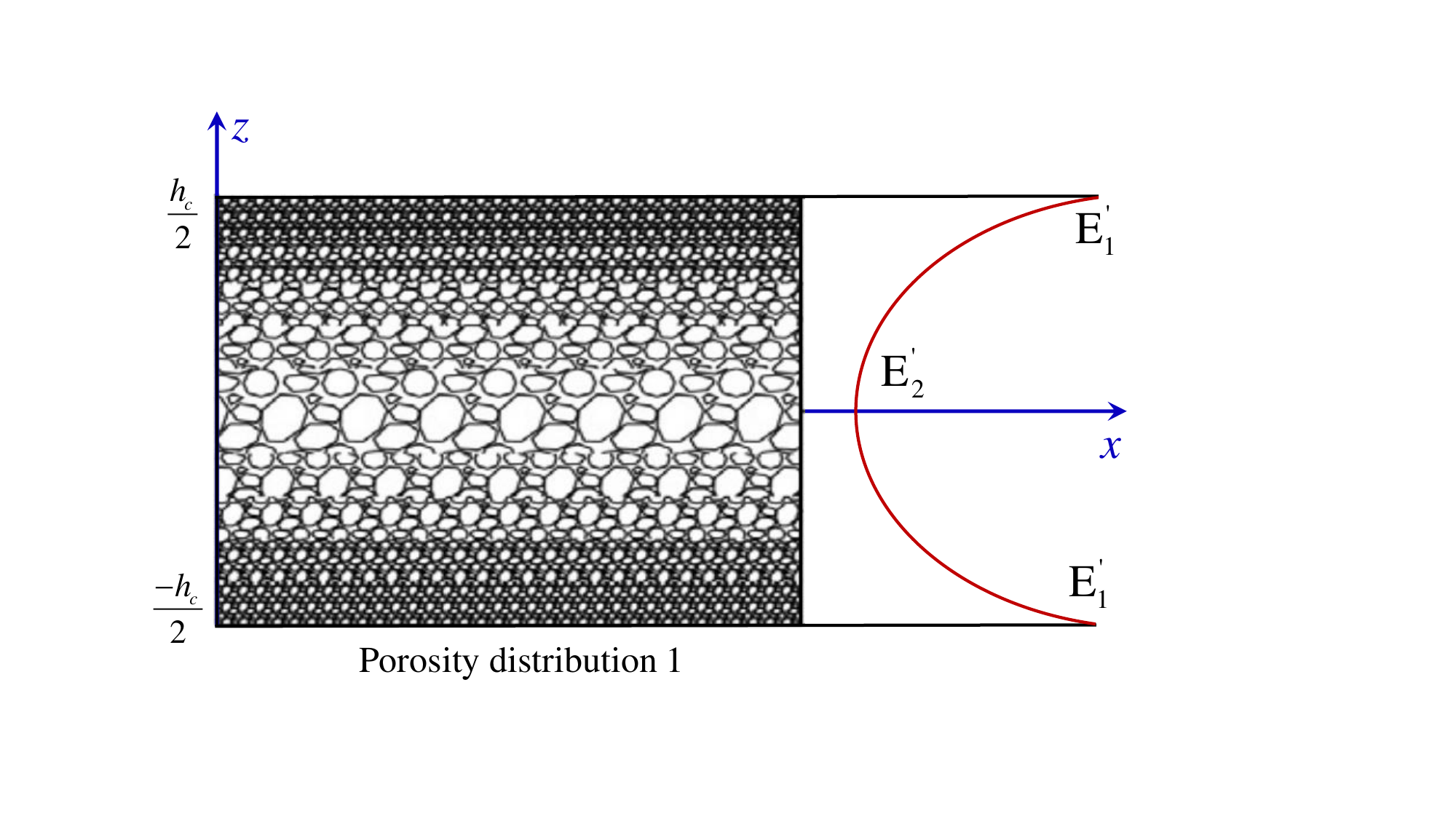}
    \end{subfigure}
    \hfill
    \begin{subfigure}[b]{0.495\textwidth}
      \centering
      \includegraphics[trim=3cm 2.5cm  7cm 2.5cm,clip=true,scale=0.3]{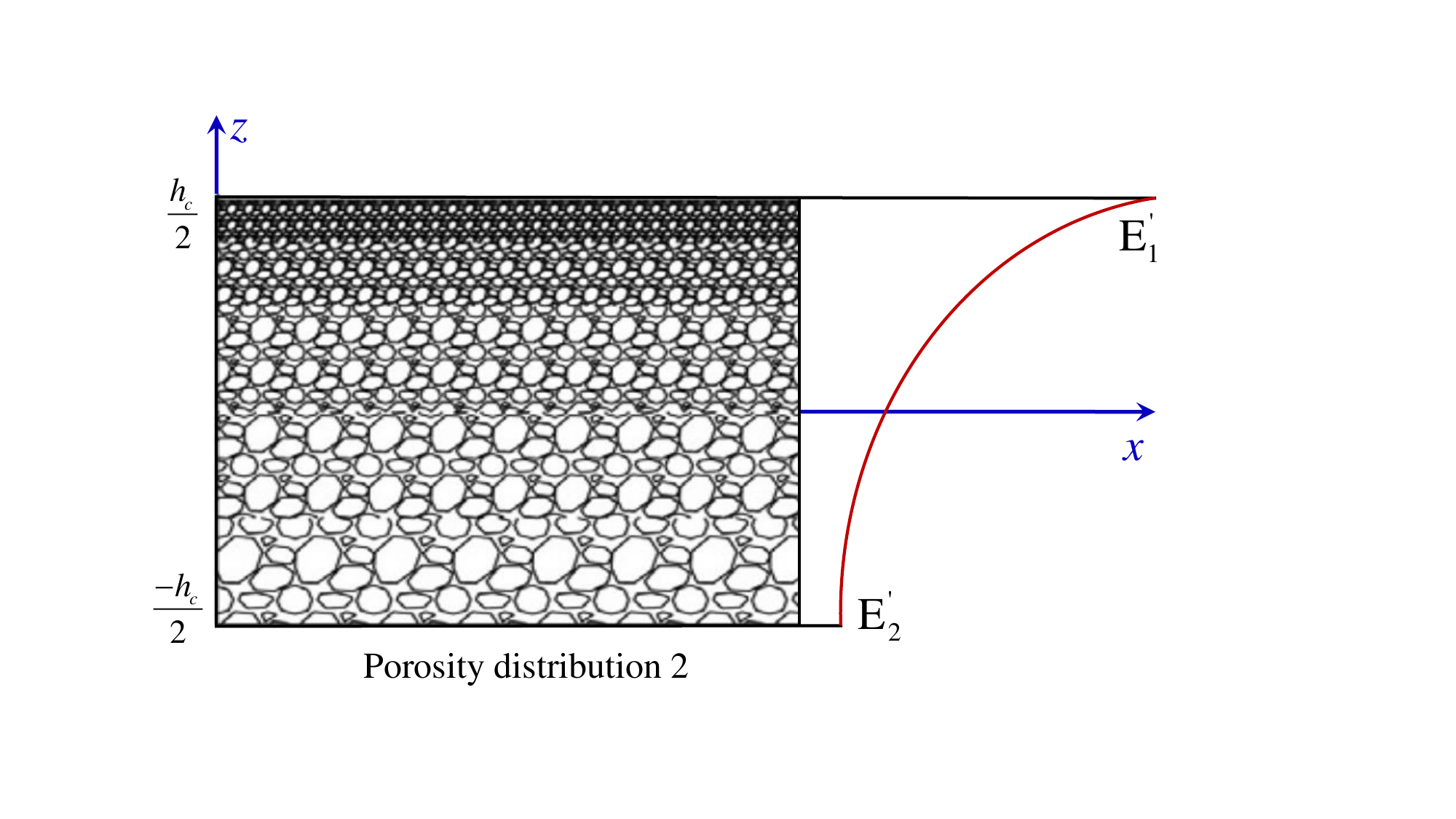}
    \end{subfigure}
    \caption{ }
    \label{fig:porous}
  \end{subfigure}
  \hfill
  \begin{subfigure}[b]{1.0\textwidth}
    \begin{subfigure}[b]{0.3\textwidth}
      \centering
      \includegraphics[trim=10cm 3.2cm 11cm 1cm,clip=true,scale=0.3]{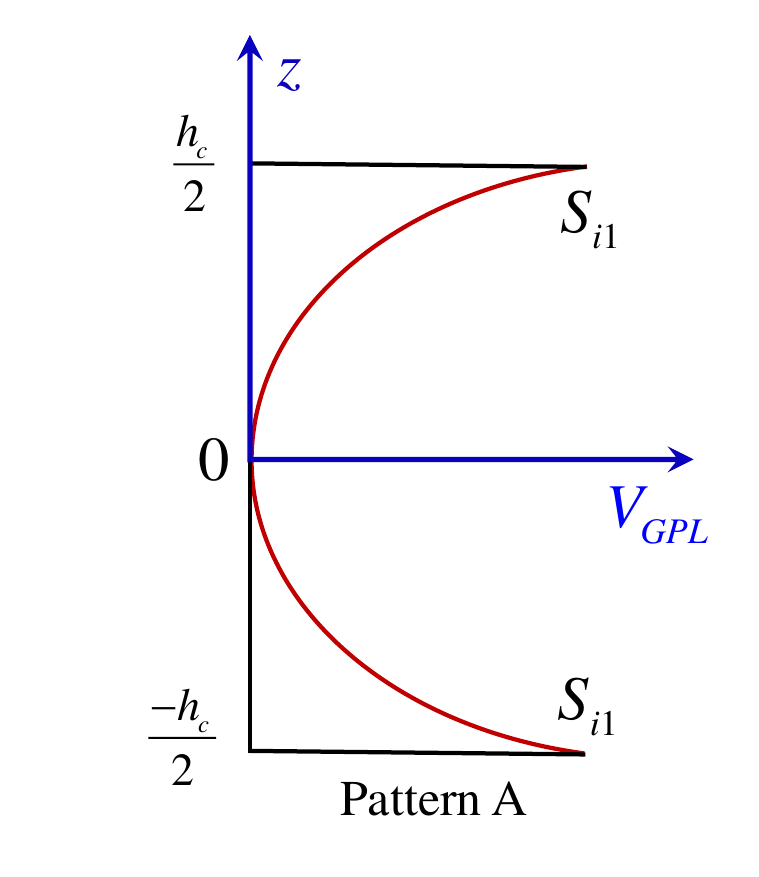}
    \end{subfigure}
    \hfill
    \begin{subfigure}[b]{0.3\textwidth}
      \centering
      \includegraphics[trim=12cm 3.2cm 11cm 0.7cm,clip=true,scale=0.3]{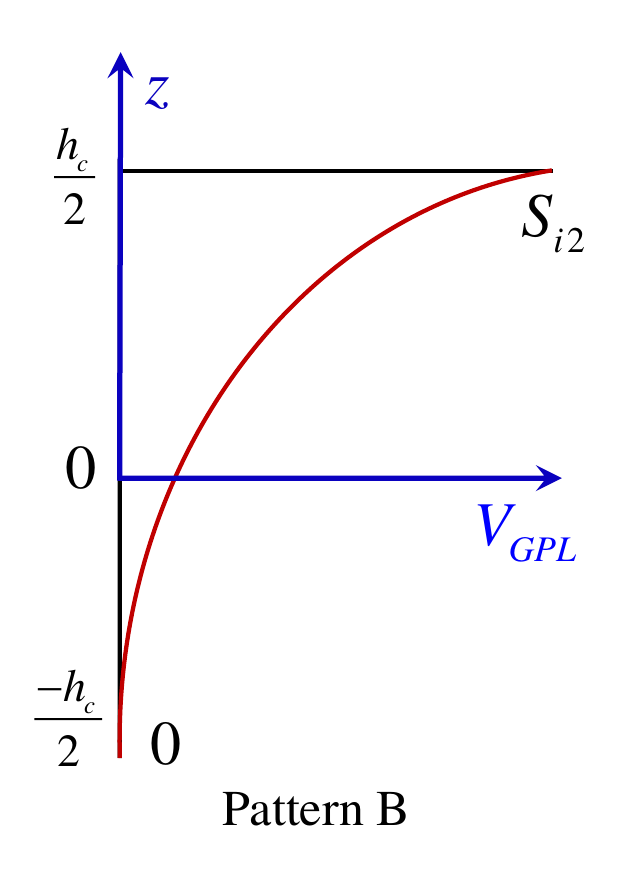}
    \end{subfigure}
    \hfill
    \begin{subfigure}[b]{0.3\textwidth}
      \centering
      \includegraphics[trim=12cm 2.2cm 11cm 0.7cm,clip=true,scale=0.3]{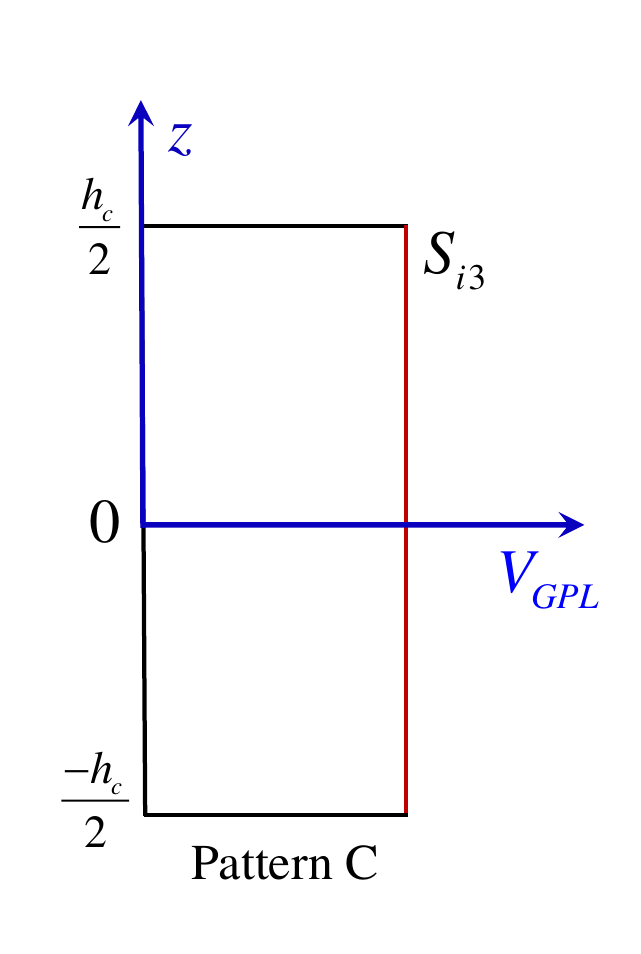}
    \end{subfigure}
    \caption{ }
    \label{fig:GPLs}
  \end{subfigure}
  \caption{Distribution of porosity and GPLs into metal matrix through the thickness of core layer $h_c$ \cite{kitipornchai2017free}: (a) distribution of porosity where $E'_1$ and $E'_2$ are the maximum and minimum values of Young's modulus without GPLs; (b) dispersion patterns of GPLs in which $S_{i1}$, $S_{i2}$ and $S_{i3}$ are the peak values of GPL volume fraction ($V_{GPL}$) with $i = 1,2$ corresponding to two porosity distributions.}
  \label{fig:porositydis}
\end{figure}

Consider a FG plate model as a sandwich structure whose core layer is porous material reinforced by GPLs and piezoelectric layers considered as face layers, as depicted in Fig. \ref{fig:Model_plates}. The geometric dimensions of FG plate are taken as length $a$, width $b$ and total thickness $h=h_c+2h_p$, where $h_c$ and $h_p$ represents the thicknesses of core layer and each piezoelectric layer, respectively. In this work, the core layers are instituted by a combination, respectively, of two porosity distributions (symmetrical and asymmetrical) and three GPL dispersion patterns (symmetrical, asymmetrical and uniform) through the thickness direction, as shown in Fig. \ref{fig:porositydis}. The effective Young's modulus $E(z)$, shear modulus $G(z)$ and mass density $\rho (z)$ for two porosity distributions can be mathematically expressed as follows
\beq
\left\{ \begin{array}{l}
  E\left( z \right) = {E_1}\left[ {1 - {e_0}\lambda \left( z \right)} \right],                      \\
  G\left( z \right) = E\left( z \right)/\left[ {2\left( {1 + \nu\left( z \right)} \right)} \right], \\
  \rho \left( z \right) = {\rho _1}\left[ {1 - {e_m}\lambda \left( z \right)} \right],
  \label{eqn:materialpro}
\end{array} \right.
\eeq

in which porosity distribution functions $\lambda (z)$ are defined as
\beq
\lambda \left( z \right) = \left\{ {\begin{array}{*{20}{l}}
      {\textrm {cos}\left( {\pi z/h_c} \right)},           & \textrm {Porosity distribution 1} \\
      {\textrm {cos}\left( {\pi z/2h_c + \pi /4} \right)}, & \textrm {Porosity distribution 2}
      \label{eqn:porositydis}
    \end{array}} \right.
\eeq
where $E_1$ and $\rho_1$ are the maximum values of Young's modulus and mass density in the thickness direction of core layer, respectively. The coefficient of porosity $e_0$ in Eq. (\ref{eqn:materialpro}) is given by
\beq
{e_0} = 1 - \frac{E'_2}{E'_1}, \;\; (0 \leq e_0 < 1),
\eeq
in which $E'_1$ and $E'_2$ denote, respectively, the maximum and minimum values of Young's modulus for core layers without GPLs reinforcement, as shown in Fig. \ref{fig:porous}. By employing the Gaussian Random Field (GRF) scheme \cite{roberts2001elastic}, the mechanical properties of closed-cell cellular solids are determined as
\beq
\frac{{E\left( z \right)}}{{{E_1}}} = {\left( {\frac{{\rho \left( z \right)/{\rho _1} + 0.121}}{{1.121}}} \right)^{2.3}}  \;  \textrm{for} \; \left( {0.15 < \frac{{\rho \left( z \right)}}{{{\rho_1}}} < 1} \right).
\label{eqn:GRF}
\eeq

From Eq. (\ref{eqn:GRF}), the mass density parameter $e_m$ in Eq. (\ref{eqn:materialpro}) are determined as
\beq
{e_m} = \frac{{1.121\left( {1 - \sqrt[{2.3}]{{1 - {e_0}\lambda \left( z \right)}}} \right)}}{{\lambda \left( z \right)}}.
\eeq

In addition, the effective Poisson's ratio $\nu(z)$ can be given as \cite{roberts2002computation}
\beq
\nu \left( z \right) = 0.221\varsigma + {\nu _1}\left( {0.342\varsigma{^2} - 1.21\varsigma+ 1} \right),
\label{eqn:Nu}
\eeq
where $\nu_1$ denotes the Poisson's ratio of metal without internal pores. The parameter $\varsigma$ in Eq. (\ref{eqn:Nu}) can be derived as follows
\beq
\varsigma= 1.121\left( {1 - \sqrt[{2.3}]{{1 - {e_0}\lambda \left( z \right)}}} \right).
\eeq

Regarding the distribution of GPLs into metal matrix, the varied volume fraction along the $z$ direction for three GPL dispersion patterns, as illustrated in Fig. \ref{fig:GPLs}, can be formulated as follows
\beq
{V_{GPL}} = \left\{ {\begin{array}{*{20}{l}}
      {{S_{i1}}\left[ {1 - \cos \left( {\pi z/h_c} \right)} \right]},           & \textrm {Pattern A} \\
      {{S_{i2}}\left[ {1 - \cos \left( {\pi z/2h_c + \pi /4} \right)} \right]}, & \textrm{Pattern B}  \\
      {{S_{i3}}},                                                               & \textrm{Pattern C}
    \end{array}} \right.
    \label{eqn:V_GPL}
\eeq
where $S_{i1}$,$S_{i2}$ and $S_{i3}$ are the peak values of GPL volume fraction, in which $i = 1,2$ corresponding to two different porosity distributions. Additionally, the relation of volume fraction and weight fraction of GPLs can be expressed as \cite{nguyen2019isogeometric}
\beq
\frac{{{\Lambda _{GPL}}{\rho _m}}}{{{\Lambda _{GPL}}{\rho _m} + {\rho _{GPL}} - {\Lambda _{GPL}}{\rho _{GPL}}}} \times \int_{ - {h_c}/2}^{{h_c}/2} {\left[ {1 - {e_m}\lambda \left( z \right)} \right]} dz = \int_{ - {h_c}/2}^{{h_c}/2} {{V_{GPL}}\left[ {1 - {e_m}\lambda \left( z \right)} \right]} dz.
\label{eqn:volu_weigh}
\eeq

From Eq. (\ref{eqn:volu_weigh}), the peak values of GPL volume fraction in Eq. (\ref{eqn:V_GPL}) can be defined. Besides, based on Halpin-Tsai micromechanics model \cite{tjong2013recent}, the value of Young's modulus $E_1$ of core layer with GPLs reinforcement without internal pores is defined as
\beq
{E_1} = \frac{3}{8}\left( {\frac{{1 + \xi _L\eta _L{V_{GPL}}}}{{1 - \eta _LV_{GPL}}}} \right){E_m} + \frac{5}{8}\left( {\frac{{1 + \xi _W\eta _W{V_{GPL}}}}{{1 - \eta _WV_{GPL}}}} \right){E_m},
\eeq
in which
\beq
\begin{array}{l}
  \xi _L = \frac{{2{l_{GPL}}}}{{{t_{GPL}}}},\;\xi _W = \frac{{2{w_{GPL}}}}{{{t_{GPL}}}},\;
  \eta _L= \frac{{\left( {{E_{GPL}}/{E_m}} \right) - 1}}{{\left( {{E_{GPL}}/{E_m}} \right) + \xi _L}},\;\eta _W = \frac{{\left( {{E_{GPL}}/{E_m}} \right) - 1}}{{\left( {{E_{GPL}}/{E_m}} \right) + \xi _W}},
\end{array}
\eeq
in which $w_{GPL}$, $l_{GPL}$ and $t_{GPL}$ are fundamental dimensions of GPLs including the average width, length and thickness, respectively. It should be mentioned that the mechanical properties of metal matrix and GPLs are symbolized with subscript symbols $m$ and $GPL$, respectively. Finally, the mass density $\rho_1$ and Poisson's ratio $\nu_1$ can be determined based on the rule of mixture \cite{nakamura2000determination}
\beq
{\rho _1} = {\rho _{GPL}}{V_{GPL}} + {\rho _m}{V_m},
\eeq
\beq
{\nu _1} = {\nu _{GPL}}{V_{GPL}} + {\nu _m}{V_m},
\eeq
in which the relation of $V_{GPL}$ and $V_{m}$ can be given as $V_{GPL} + V_m = 1$.

\subsection{$C^0$-type higher-order shear deformation theory}
\label{sec:plateTheory}
Considering a plate carrying a domain ${\bf V}=\Omega\times(\frac{-h}{2},\frac{h}{2})$, in which $\Omega\in \mathbb{R}^2$. Assuming that the displacement components of plate in the $x$, $y$, $z$ directions and rotations
about the $y$- and the $x$- axes are $u_0, v_0, w_0, \theta_x $ and $\theta_y$, respectively. Based on the higher-order shear deformation theory \cite{aydogdu2009new}, the displacement field at an any point in plate can be described as follows
\beq
{\bf{u}}(x,y,z) = {\bf{u}}^0(x,y) + z{\bf{u}}^1(x,y) + f(z){\bf{u}}^2(x,y),
\label{eqn:dispstrain}
\eeq
where
\beq
{\bf{u}} = \left\{ {\begin{array}{*{20}{c}}
      u \\v\\w
    \end{array}} \right\},{\rm{  }}{{\bf{u}}^0} = \left\{ {\begin{array}{*{20}{c}}
      {{u_0}}\\{{v_0}}\\{{w_0}}
    \end{array}} \right\},\,{\rm{  }}\,{{\bf{u}}^1} =  - \left\{ {\begin{array}{*{20}{c}}
      {{w_{0,x}}}\\{{w_{0,y}}} \\0
    \end{array}} \right\},{\rm{  }}{{\bf{u}}^2} = \left\{ {\begin{array}{*{20}{c}}
      {{\theta _x}}\\{{\theta _y}} \\0
    \end{array}} \right\},
\label{eqn:dispstrain1}
\eeq
where the subscript symbols $x$ and $y$ denote the derivative of any function corresponding to $x$ and $y$ directions, respectively. Meanwhile, the function $f(z)$ describes the shear strain and stress distribution along the thickness, as found in \cite{nguyen2016general}. In this work, the famous third-order function proposed by Reddy is utilized as $f(z)=z-\frac{4z^3}{3h^2}$ \cite{reddy2000analysis}.\\

In order to avoid the high-order derivations in approximate formulation and enforce boundary conditions conveniently, two extra assumptions are formulated as
\beq
w_{0,x}=\beta_x,\;w_{0,y}=\beta_y.
\label{eqn:dispstrain2}
\eeq
By substituting Eq. (\ref{eqn:dispstrain2}) into Eq. (\ref{eqn:dispstrain1}), yields
\beq
{{\bf{u}}^0} = {\left\{ {\begin{array}{*{20}{c}}
      {{u_0}}\\{{v_0}}\\{{w_0}}
    \end{array}} \right\}},{{\bf{u}}^1} =  - {\left\{ {\begin{array}{*{20}{c}}
      {{\beta _x}}\\{{\beta _y}} \\0
    \end{array}} \right\}},{{\bf{u}}^2} = {\left\{ {\begin{array}{*{20}{c}}
          {{\theta _x}}\\{{\theta _y}} \\0
        \end{array}} \right\}.}
\label{eqn:dispstrain3}
\eeq

It can be seen that the strain fields in Eq. (\ref{eqn:dispstrain3}) only involve the $C^0$-continuity of approximate field. The Green strain vector of a bending plate can be written in compact form as follows
\beq
{\varepsilon _{ij}} = \frac{1}{2}\left( {\frac{{\partial {u_i}}}{{\partial {x_j}}} + \frac{{\partial {u_j}}}{{\partial {x_i}}} + \frac{{\partial {u_k}}}{{\partial {x_i}}}\frac{{\partial {u_k}}}{{\partial {x_j}}}} \right).
\label{eqn:Green_strain}
\eeq

By employing the von K\'arm\'an assumptions, the relationship of strain and displacement is given as
\bseq
\begin{align}
  \vvarepsilon={\left\{ {{\vvarepsilon _{xx}},{\vvarepsilon _{yy}},{\vgamma _{xy}}} \right\}^T} = {{\vvarepsilon} _m} + z{{\vkappa }_1} + {f(z)}{{\vkappa }_2}, \\
  {\bf{\vgamma }} = {\left\{ {{\gamma _{xz}},{\gamma _{yz}}} \right\}^T} = {{\bf{\vvarepsilon }}_s} + {f'(z)}{{\bf{\vkappa }}_s},
  \label{eqn:strainplane}
\end{align}
\eseq
where
\beq
\begin{array}{l}
  {{\vvarepsilon }_m} = \left\{ {\begin{array}{*{20}{c}}
        {{u_{0,x}}} \\
        {{v_{0,y}}} \\
        {{u_{0,y}} + {v_{0,x}}}
      \end{array}} \right\} + \frac{1}{2}\left\{ {\begin{array}{*{20}{c}}
        {w_{0,x}^2} \\
        {w_{0,y}^2} \\
        {2w_{0,xy}^{}}
      \end{array}} \right\} = {\vvarepsilon }_m^L + {\vvarepsilon }_m^{NL}, \\
  {{\vkappa }_1} =  - \left\{ {\begin{array}{*{20}{c}}
        {{\beta _{x,x}}} \\
        {{\beta _{y,y}}} \\
        {{\beta _{x,y}} + {\beta _{y,x}}}
      \end{array}} \right\},{{\vkappa }_2} = \left\{ {\begin{array}{*{20}{c}}
        {{\theta _{x,x}}} \\
        {{\theta _{y,y}}} \\
        {{\theta _{x,y}} + {\theta _{y,x}}}
      \end{array}} \right\},
  {\vvarepsilon _{s}} = \left\{ {\begin{array}{*{20}{c}}
        {{w_{0,x}} - {\beta _x}} \\
        {{w_{0,y}} - {\beta _y}}
      \end{array}} \right\},{\vkappa_{s}} = \left\{ {\begin{array}{*{20}{c}}
        {{\theta _x}} \\
        {{\theta _y}}
      \end{array}} \right\}
\end{array}
\label{eqn:extract_disp}
\eeq
where the nonlinear strain component is expressed as follows
\beq
{\vvarepsilon }_m^{NL}=\frac{1}{2}\left[ {\begin{array}{*{20}{c}}
        {{w_{0,x}}} & 0          \\
        0          & {{w_{0,y}}} \\
        {{w_{0,y}}} & {{w_{0,x}}}
      \end{array}} \right]\left\{ {\begin{array}{*{20}{c}}
      {{w_{0,x}}} \\
      {{w_{0,y}}}
    \end{array}} \right\} = \frac{1}{2}{\boldsymbol\Theta} {\bf{\Lambda }}.
\eeq

\subsection{Constitutive equations}
In this study, the constitutive relationship of FG plate embedded in piezoelectric layers is expressed as \cite{wang2001vibration}
\beq
\left\{ {\begin{array}{*{20}{c}}
        {\boldsymbol{\sigma }} \\
        {\bf{D}}
      \end{array}} \right\} = \left[ {\begin{array}{*{20}{c}}
        {\bf{c}} & { - {{\bf{e}}^T}} \\
        {\bf{e}} & {\bf{g}}
      \end{array}} \right]\left\{ {\begin{array}{*{20}{c}}
        {\hat {\boldsymbol\varepsilon} } \\
        {\bf{E}}
      \end{array}} \right\},
      \label{eqn:consti}
\eeq
in which ${\boldsymbol{\sigma }}$ and $\hat {\boldsymbol\varepsilon}  = {{\rm{\{}}{{\boldsymbol{\varepsilon }}},\;{{\boldsymbol{\gamma}}}{\rm{\}}}^T}$ represent, respectively, the stress and strain vectors of mechanical field. $\bf D$ stands for dielectric displacement vector; Meanwhile, $\bf c$, $\bf e$ and $\bf g$ are the mechanical, piezoelectric and dielectric constant matrices, respectively. In addition, the electric field vector $\bf E$ is calculated based on the electric potential field $\phi$ as follows \cite{tzou1990distributed}
\beq
{\bf{E}} =  - {\rm{grad}}\phi =  - \nabla \phi.
\label{eqn: elec_potent}
\eeq
\\
in which $\phi$ stands for the electric potential difference through the piezoelectric layer. The material constant matrix $\bf c$ can be expressed as
\beq
{\bf{c}} = \left[ {\begin{array}{*{20}{c}}
        {\bf{A}} & {\bf{B}} & {\bf{N}} & {\bf{0}}       & {\bf{0}}       \\
        {\bf{B}} & {\bf{M}} & {\bf{G}} & {\bf{0}}       & {\bf{0}}       \\
        {\bf{N}} & {\bf{G}} & {\bf{H}} & {\bf{0}}       & {\bf{0}}       \\
        {\bf{0}} & {\bf{0}} & {\bf{0}} & {{{\bf{A}}^s}} & {{{\bf{B}}^s}} \\
        {\bf{0}} & {\bf{0}} & {\bf{0}} & {{{\bf{B}}^s}} & {{{\bf{D}}^s}}
      \end{array}} \right],
\eeq
in which
\beq
\begin{array}{l}
  \left( {\bf A,\bf B,\bf M,\bf N,\bf G,\bf H} \right) = \int_{ - h_c/2}^{h_c/2} {\left( {1,z,{z^2},f(z),zf(z),{f^2}(z)} \right)}Q^bdz, \\
  \left( {{\bf A}_{}^s,{\bf B}_{}^s,{\bf D}_{}^s} \right) = \int_{ - h_c/2}^{h_c/2} {(1,f'(z),{{f'}^2}(z))}Q^sdz,
\end{array}
\label{eqn:matmatrices1}
\eeq
in which
\beq
{\bf Q}^b = \frac{{{E(z)}}}{{1 - \nu(z)^2}}\left[ {\begin{array}{*{20}{c}}
        1          & {{\nu (z)}} & 0                       \\
        {{\nu(z)}} & 1          & 0                       \\
        0          & 0          & {\frac {1 - \nu (z)}{2}}
      \end{array}} \right],\;\;{\bf Q}^s  = \frac{{{E(z)}}}{{2(1 + \nu (z)^{})}}\left[ {\begin{array}{*{20}{c}}
        1 & 0 \\
        0 & 1
      \end{array}} \right].
\eeq
where $E(z)$ and $\nu_(z)$ are the effective Young's modulus and Poisson's ratio, respectively. Additionally, the constant matrices of piezoelectric material, as presented in Eq. (\ref{eqn:consti}), including the stress piezoelectric $\bf e$ and the dielectric $\bf g$ are given by \cite{wang2004finite}
\beq
{\bf{e}} = \left[ {\begin{array}{*{20}{c}}
  0          & 0          & 0          & 0          & {{e_{15}}} \\
  0          & 0          & 0          & {{e_{15}}} & 0          \\
  {{e_{31}}} & {{e_{32}}} & {{e_{33}}} & 0          & 0
\end{array}{\rm{  }}\begin{array}{*{20}{c}}
\end{array}} \right],{\bf{g}} = \left[ {\begin{array}{*{20}{c}}
        {{p_{11}}} & 0          & 0          \\
        0          & {{p_{22}}} & 0          \\
        0          & 0          & {{p_{33}}}
      \end{array}} \right].
\eeq

\subsection{Weak form of governing equations}

The governing equations of motion for piezoelectric FG plate can obtain by applying the Hamilton's variational principle which are given as follows \cite{hwang1993finite}
\beq
\delta \int_{{t_1}}^{{t_2}} {Ldt = 0},
\label{eqn:galerkin}
\eeq
in which $t_1$ and $t_2$ represent the starting and end time values, respectively. The general energy function $L$ is the summation of kinetic energy, strain energy, dielectric energy and external work as follows
\beq
L = \frac{1}{2}\int_\Omega  {\left( {\rho {{{\bf{\dot u}}}^T}{\bf{\dot u}} - {{\bf{\vsigma }}^T}\vvarepsilon  + {{\bf{D}}^T}{\bf{E}}} \right)} {\rm{d}}\Omega  + \int_{{\Gamma _s}} {{{\bf{u}}^T}{{\bf{f}}_s}} {\rm{d}}{\Gamma _s} - \int_{{\Gamma _\phi }} {{{\phi} }{{\bf{q}}_s}} {\rm{d}}{\Gamma _\phi } + \sum {{{\bf{u}}^T}{{\bf{F}}_p}}  - \sum {{\phi} {{\bf{Q}}_p}},
\label{eqn:total_Ener}
\eeq
where $\rho$ represents the mass density; $\bf u$ and $\dot{\bf u}$ denote the mechanical displacement and velocity vectors, respectively;  ${\bf f}_s$  and ${\bf F}_p$ are, respectively, the external mechanical surface and concentrated load vectors while ${\bf q}_s$ and ${\bf Q}_p$ are the external surface and point charges, respectively; $\Gamma_s$ and $\Gamma_\phi$ indicate the external mechanical and the electrical loading surface, respectively.\\

Then, the variational form of Eq. (\ref{eqn:total_Ener}) can be expressed as follows
\beq
\begin{array}{l}
  \int_{{t_1}}^{{t_2}} {\int_\Omega  {\left( {\rho {\bf{\ddot u}}\delta {\bf{u}}^T - {\bf{\vsigma }}^T\delta {\boldsymbol\varepsilon}  + {\bf{D}}^T\delta {\bf{E}}} \right)} } {\rm{d}}\Omega {\rm{dt + }}\int_{{t_1}}^{{t_2}} {\int_{{\Gamma _s}} {{{\bf{f}}_s}\delta {\bf{u}}^T} } {\rm{d}}{\Gamma _s}{\rm{dt - }}\int_{{t_1}}^{{t_2}} {\int_{{\Gamma _\phi }} {{{\bf{q}}_s}\delta {\phi} } } {\rm{d}}{\Gamma _\phi }{\rm{dt}} \\
  \\
  {\rm{ + }}\int_{{t_1}}^{{t_2}} {\sum {{{\bf{F}}_p}} } \delta {{\bf{u}}^T}dt - \int_{{t_1}}^{{t_2}} {\sum {{{\bf{Q}}_p}} } \delta \phi dt = 0.
\end{array}
\label{eqn:varia_form}
\eeq

\section{Isogeometric analysis based on B\'ezier extraction}
\label{sec:IGA_Section}
\subsection{B-spline and NURBS basis functions}

In the parametric space for one dimensional (1D) problem, B-spline basis functions can be established by  non-decreasing set real-valued coordinates called knot vector ${\bf{\Xi }} = \left\{ {{\xi _1},{\xi _2},...,{\xi _{n + p + 1}}} \right\}$, in which $\xi_i$ is the $i$th knot with $i = 1,...n + p+1$ indicates the knot index. Meanwhile, $n$ and $p$ are the number of univariate B-spline basis functions and its order, respectively. For a given knot vector $\bf{\Xi }$, the B-spline basis functions are defined according to recursive form \cite{hughes2005isogeometric, cottrell2009isogeometric}
\beq
{N_{i,0}}\left( \xi  \right) = \left\{ {\begin{array}{*{20}{l}}
      1, & \textrm{ if}\;\;{{\xi _i} \le \xi  \le {\xi _{i + 1}}} \\
      0, & \textrm {otherwise}
    \end{array}} \textrm {for} \;\;\; p=0, \right.
\eeq

\beq
{N_{i,p}}\left( \xi  \right) = \frac{{\xi  - {\xi _i}}}{{{\xi _{i + p}} - {\xi _i}}}{N_{i,p - 1}}\left( \xi  \right) + \frac{{{\xi _{i + p + 1}} - \xi }}{{{\xi _{i + p + 1}} - {\xi _{i + 1}}}}{N_{i + 1,p - 1}}\left( \xi  \right) \textrm {for} \;\;\; p>0.
\eeq

Then, B-spline curves can be constructed by taking a linear combination of B-spline basis functions and control points ${P_i}\left( {i = 1,2,...,n} \right)$ as
\beq
{\bf C}\left( \xi  \right) = \sum\limits_{i=1}^n {{\bf P}_i{N_{i,p}}\left( \xi  \right)}.
\eeq
\\
The B-spline basis functions in two dimensional (2D) space can also obtain by taking a tensor product of two 1D B-splines basis functions. The B-splines surfaces are then expressed as follows \cite{hughes2005isogeometric}
\beq
{\bf S}\left( {\xi ,\eta } \right) = \sum\limits_{i = 1}^n {\sum\limits_{j = 1}^m {{\bf P}_{i,j}{N_{i,p}}} } \left( \xi  \right){M_{j,q}}\left( \eta  \right) = {\bf P}^T{\bf N}\left( {\xi ,\eta } \right),
\label{eqn:B-spli_sur}
\eeq
in which ${N_{i,p}}$ and ${M_{j,q}}$ represent the 1D B-splines basis functions having orders $p$ and $q$ in the $\xi$ and $\eta$ directions corresponding with two knot vectors ${\bf{\Xi }} = \left\{ {{\xi _1},{\xi _2},...,{\xi _{n + p + 1}}} \right\}$ and ${\boldsymbol{\Theta}} = \left\{ {{\eta _1},{\eta _2},...,{\eta _{m + q + 1}}} \right\}$, respectively.\\

In order to describe exactly some conic shapes such as circle, cylinder, sphere and ellipsoid, the NURBS is introduced based on the B-spline and a set of weights \cite{cottrell2009isogeometric}. Accordingly, the NURBS basis functions can be expressed as
\beq
{R_{i,j}}\left( {\xi ,\eta } \right) = \frac{{{N_{i,p}}\left( \xi  \right){M_{j,q}}\left( \eta  \right){w_{i,j}}}}{{\sum\limits_{{\hat i}=1}^n {\sum\limits_{{\hat j}=1}^m {{N_{{{\hat i}},p}}\left( \xi  \right){M_{{{\hat j}},q}}\left( \eta  \right){w_{\hat i,\hat j}}}}}},
\eeq
where ${w_{i,j}}$ represents the weight values. Finally, the NURBS surfaces can be determined as follows
\beq
{\bf S}\left( {\xi ,\eta } \right) = \sum\limits_{i = 1}^n {\sum\limits_{j = 1}^m {{R_{i,j}}\left( {\xi ,\eta } \right)} } {{\bf P}_{i,j}}.
\eeq

\subsection{B\'ezier extraction of NURBS}
The main aim of B\'ezier extraction technique is to instead the NURBS basis functions by the $C^0$-continuous Bernstein polynomial basis functions described over B\'ezier elements which have the similar element structure with standard FEM. As a result, the IGA is straightforwardly performed as well as can be integrated into most available FEM framework. It is well known that the B-spline basis functions of $p$th order possess $C^{p-k}$ continuity across each interior knots, in which $k$ represents the multiplicity of knots in the knot value. Hence, we can obtain the $C^0$-continuity by inserting the new knots into the original knot vector until $k$ is equal to $p$. By inserting a new knot $\bar \xi  \in \left[ {{\xi _k},{\xi _{k + 1}}} \right]$ with $(k>p)$ into the original knot vector ${\bf{\Xi }} = \left\{ {{\xi _1},{\xi _2},...,{\xi _{n + p + 1}}} \right\}$, a new set of control points ${\bar {\bf P}_i}$ are found and expressed as follows \cite{hughes2005isogeometric}
\beq
{\bar {\bf P}_i} = \left\{ {\begin{array}{*{20}{l}}
      {{{\bf {P}}_1}}                                                        \\
      {{\alpha _i}{{\bf {P}}_i}{\rm{ + (1 - }}{\alpha _i}){{\bf {P}}_{i - 1}}} \\
      {{{\bf {P}}_n}}
    \end{array}} \right.{\rm{ }}\begin{array}{*{20}{l}}
  {{\rm{ }}i = 1}, \\
  {1 < i < n + 1}, \\
  {i = n + 1},
\end{array}
\eeq
in which
\beq
{\alpha _i} = \left\{ {\begin{array}{*{20}{l}}
      1                                                          & 1\le  i \le k - p,       \\
      \frac{{\bar \xi  - {\xi _i}}}{{{\xi _{i + p}} - {\xi _i}}} & {k - p + 1 \le i \le k}, \\
      0                                                          & {i \ge k + 1},
    \end{array}} \right.
\eeq

Then, the B\'ezier extraction operator can be established by using the new set of knots as follows \cite{borden2011isogeometric, do2017limit}
\beq
{\bf C}^j = \left[ {\begin{array}{*{20}{c}}
        {{\alpha _1}} & {1 - {\alpha _2}} & 0                 & \ldots            & {}     & {}                      & 0                       \\
        0             & {{\alpha _2}}     & {1 - {\alpha _3}} & 0                 & \ldots & {}                      & 0                       \\
        0             & 0                 & {{\alpha _3}}     & {1 - {\alpha _4}} & 0      & \ldots                  & 0                       \\
        \vdots        & {}                & {}                & {}                & {}     & {}                      & {}                      \\
        0             & \ldots            & {}                & {}                & 0      & {{\alpha _{n + j - 1}}} & {1 - {\alpha _{n + j}}}
      \end{array}} \right].
\eeq

By employing the B\'ezier extraction operator ${\bf C}^j$, a new B\'ezier control points ${\bf P}^b$ associated with Bernstein polynomial basis can obtain as follows \cite{thomas2015bezier}
\beq
{\bf P}^b = {\bf C}^T\bf P,
\label{eqn:new_cont_point}
\eeq
in which the whole B\'ezier extraction operator $\bf C$ can be given as
\beq
{\bf C} = {\prod\nolimits_{j = 1}^m {{{\bf C}^j}}}.
\eeq

It should be noted that the above insertion into the original knot vector does not lead to any change of the geometry. Consequently, the B-spline surface can also obtain by using B\'ezier control points and Bernstein polynomials as
\beq
{\bf S}\left( {\xi ,\eta } \right) = \sum\limits_i^n {\sum\limits_j^m {{B_{i,j}}\left( {\xi ,\eta } \right){\bf P}_{i,j}^b} }  = {\left( {{\bf P}^b} \right)^T}{\bf B}\left( {\xi ,\eta } \right),
\label{eqn:Bezier_sur}
\eeq
where 2D Bernstein polynomials ${\bf B}\left( {\xi ,\eta } \right)$ are defined recursively as
\beq
\begin{array}{l}
  {B_{i,j,p}}\left( {\xi ,\eta } \right) = \frac{1}{4}\left( {1 - \xi } \right)\left( {1 + \eta } \right){B_{i,j - 1,p - 1}}\left( {\xi ,\eta } \right) + \frac{1}{4}\left( {1 - \xi } \right)\left( {1 - \eta } \right){B_{i,j,p - 1}}\left( {\xi ,\eta } \right) + \\
  \frac{1}{4}\left( {1 + \xi } \right)\left( {1 - \eta } \right){B_{i - 1,j,p - 1}}\left( {\xi ,\eta } \right) + \frac{1}{4}\left( {1 + \xi } \right)\left( {1 + \eta } \right){B_{i - 1,j - 1,p - 1}}\left( {\xi ,\eta } \right),
\end{array}
\eeq
in which
\beq
{B_{1,1,0}}\left( {\xi ,\eta } \right) = 1,\;{\rm{ }}{B_{i,j,p}}\left( {\xi ,\eta } \right) = 0\; \left( {i,j < 1\; \textrm{ or }\; i{\rm{, }}j{\rm{ > {p+1}}}} \right).
\eeq
Comparing Eqs. (\ref{eqn:B-spli_sur}) and (\ref{eqn:Bezier_sur}), yields
\beq
{\bf P}^T{\bf N}\left( {\xi ,\eta } \right)={\left( {{\bf P}^b} \right)^T}{\bf B}\left( {\xi ,\eta } \right).
\label{eqn:relation_2B}
\eeq

By substituting  Eq. (\ref{eqn:new_cont_point}) into Eq. (\ref{eqn:relation_2B}), the relationship between the B-spline and Bernstein polynomial functions can obtain as follows
\beq
{\bf N}\left( {\xi ,\eta } \right) = {\bf CB}\left( {\xi ,\eta } \right),
\label{eqn:new_Nurbs}
\eeq

Based on Eq. (\ref{eqn:new_Nurbs}), the NURBS basis functions can be expressed by Bernstein polynomials as follows
\beq
{\bf R}\left( {\xi ,\eta } \right) = \frac{\bf W}{{W\left( {\xi ,\eta } \right)}}{\bf N}\left( {\xi ,\eta } \right) = \frac{\bf W}{{W\left( {\xi ,\eta } \right)}}{\bf CB}\left( {\xi ,\eta } \right),
\label{eqn: NURBS_basic_func}
\eeq
where $\bf{W}$ denotes the local NURBS weights in diagonal matrix. Meanwhile, the weight functions $W\left( {\xi ,\eta } \right)$ are described based on Bernstein basis functions as follows
\beq
W\left( {\xi ,\eta } \right) = {\left( {{\bf C}^T{\bf w}} \right)^T}{\bf B}\left( {\xi ,\eta } \right) = {\left( {{\bf w}^b} \right)^T}{\bf B}\left( {\xi ,\eta } \right),
\eeq
in which $\bf{w}$ and $\bf{w}^b$ indicate, respectively the weights of the NURBS and B\'ezier. In addition, the relationship of B\'ezier and NURBS control points can be given as
\beq
{\bf P}^b = {\left( {{\bf W}^b} \right)^{ - 1}}{\bf C}^T{\bf {WP}}.
\eeq

\subsection {Discrete system equations}
\subsubsection {Mechanical displacement field}

By employing B\'ezier extraction of NURBS, the approximation of mechanical displacement field $\bf{u{(\xi ,\eta) }}$ for FG plate is expressed as
\beq
{\bf{u}}^h\left( {\xi ,\eta } \right) = \sum\limits_{I=1}^{m \times n} {R_I^e\left( {\xi ,\eta } \right)} \bf{d_I},
\label{eqn:appx_mech_disp}
\eeq
in which $m\times n$ represents the number of basis functions and ${R}_I^e\left( {\xi ,\eta } \right)$ denotes the NURBS basis functions which are presented in Eq. (\ref{eqn: NURBS_basic_func}). The vector ${\bf d}_I = {\left\{ {\begin{array}{*{20}{c}}{{u_{0I}}}&{{v_{0I}}}&{{w_{0I}}}&{{\beta _{xI}}}&{{\beta _{yI}}}&{{\theta _{xI}}}&{{\theta _{yI}}} \end{array}} \right\}^T}$ includes the nodal degrees of freedom associated with control point I.\\

Substituting Eq. (\ref{eqn:appx_mech_disp}) into Eq. (\ref{eqn:extract_disp}), the strain components are expressed as
\beq
{\left\{ {{\boldsymbol{\varepsilon }},{\boldsymbol{\gamma }}} \right\}^T} = \sum\limits_{I = 1}^{m \times n} {\left( {{\bf{B}}_I^L + \frac{1}{2}{\bf{B}}_I^{NL}} \right)} {{\bf{d}}_I},
\eeq
in which ${\bf{B}}_I^L = {\left[ {\begin{array}{*{20}{c}} {{\bf{B}}_I^m}&{{\bf{B}}_I^{b1}}&{{\bf{B}}_I^{b2}}&{{\bf{B}}_I^{s1}}&{{\bf{B}}_I^{s2}} \end{array}} \right]^T}$ is given as follows
\beq
\begin{array}{l}
  {\bf B}^m = {\left[ {\begin{array}{*{20}{c}}
          {{R_{I,x}}} & 0           & 0 & 0 & 0 & 0 & 0 \\
          0           & {{R_{I,y}}} & 0 & 0 & 0 & 0 & 0 \\
          {{R_{I,y}}} & {{R_{I,x}}} & 0 & 0 & 0 & 0 & 0
        \end{array}} \right]},\;{\bf B}^{b1} =  - {\left[ {\begin{array}{*{20}{c}}
          0 & 0 & 0 & {{R_{I,x}}} & 0           & 0 & 0 \\
          0 & 0 & 0 & 0           & {{R_{I,y}}} & 0 & 0 \\
          0 & 0 & 0 & {{R_{I,y}}} & {{R_{I,x}}} & 0 & 0
        \end{array}} \right]}, \\
  \\
  {\bf B}^{b2} =  {\left[ {\begin{array}{*{20}{c}}
          0 & 0 & 0 & 0 & 0 & {{R_{I,x}}} & 0           \\
          0 & 0 & 0 & 0 & 0 & 0           & {{R_{I,y}}} \\
          0 & 0 & 0 & 0 & 0 & {{R_{I,y}}} & {{R_{I,x}}}
        \end{array}} \right]},                                                               \\
  \\
  {\bf B}^{s1} = {\left[ {\begin{array}{*{20}{c}}
              0 & 0 & {{R_{I,x}}} & { - {R_I}} & 0          & 0 & 0 \\
              0 & 0 & {{R_{I,y}}} & 0          & { - {R_I}} & 0 & 0
            \end{array}} \right]},\;{\bf B}^{s2} =  {\left[ {\begin{array}{*{20}{c}}
              0 & 0 & 0 & 0 & 0 & {{R_I}} & 0       \\
              0 & 0 & 0 & 0 & 0 & 0       & {{R_I}}\end{array}} \right]},
\end{array}
\eeq
and ${\bf{B}}_I^{NL} =  {\boldsymbol\Theta} {\boldsymbol\Lambda}$ with
\beq
{\boldsymbol\Theta}  =  {\left[ {\begin{array}{*{20}{c}}
        w_{I,x} & 0       \\
        0       & w_{I,y} \\
        w_{I,y} & w_{I,x} \\
      \end{array}} \right]},\;{\boldsymbol\Lambda}= {\left[ {\begin{array}{*{20}{c}}
        0 & 0 & R_{I,x} & 0 & 0 & 0 & 0 \\
        0 & 0 & R_{I,y} & 0 & 0 & 0 & 0
      \end{array}} \right]}.
\eeq

\subsubsection{Electric potential field}

The electric potential field on piezoelectric layers is estimated by dividing each layer into $n_{sub}$ sublayers through the thickness. In each sublayer, the electric potential variation is assumed to be linear across thickness and calculated as follows \cite{wang2004finite}
\beq
{\phi ^h}(z) = {\bf R}_\phi {\boldsymbol\phi}_I ,
\eeq
where ${\bf R}_\phi $ indicates the shape function of the electric potential function which is determined in Eq. (\ref{eqn: NURBS_basic_func}) with order $p = 1$. The function ${\boldsymbol\phi}_I  = \left\{ {\begin{array}{*{20}{c}}{{\phi ^{i - 1}}}, & {{\phi ^i}}\end{array}} \right\}$ with $i = 1,2,...,{n_{sub}}$ is the electric potentials at the upper and lower surfaces of each sublayer.\\

Assuming that at the same height, the values of electric potentials are set to be equal \cite{wang2001vibration}. As a result, the electric potential field $\bf E$ for each sublayer element can be given as
\beq
{\bf{E}} =  - \nabla {\bf R}_\phi {\boldsymbol\phi}_I =  - {\bf{B}}_\phi ^{}{\boldsymbol\phi}_I,
\label{eqn:approx_elec}
\eeq\\
where
\beq
{\bf{B}}_\phi ^{} = \left\{ {\begin{array}{*{20}{c}}
      0 & 0 & {\frac{1}{{{h_p}}}}
    \end{array}} \right\}^T.
\eeq

\subsection{Governing equation of motion}
By substituting Eqs. (\ref{eqn:appx_mech_disp}) and (\ref{eqn:approx_elec}) into Eq. (\ref{eqn:varia_form}), the final form of governing equation of motion can obtain as follows \cite{wang2004finite}
\beq
\left[ {\begin{array}{*{20}{c}}
        {{{\bf{M}}_{uu}}} & {\bf 0} \\
        {\bf 0}           & {\bf 0}
      \end{array}} \right]\left[ {\begin{array}{*{20}{c}}
        {{\ddot{\bf d}}} \\
        {\ddot{\boldsymbol\phi}}
      \end{array}} \right] +
\left[ {\begin{array}{*{20}{c}}
        {{{\bf{K}}_{uu}}}     & {{{\bf{K}}_{u\phi }}}      \\
        {{{\bf{K}}_{\phi u}}} & {{{-\bf{K}}_{\phi \phi }}}
      \end{array}} \right]\left[ {\begin{array}{*{20}{c}}
        {\bf{d}} \\
        {\boldsymbol\phi}
      \end{array}} \right] = \left[ {\begin{array}{*{20}{c}}
        {\bf{F}} \\
        {\bf{Q}}
      \end{array}} \right],
\label{eqn:General_FEM}
\eeq
where

\beq
\begin{array}{l}
  {{\bf{K}}_{uu}} = \int_\Omega  {({{\bf B}^L+{\bf B}^{NL})}^T{\bf c}({{\bf B}^L+\frac{1}{2}{\bf B}^{NL})}d\Omega },\;{{\bf{K}}_{\phi \phi }} = \int_\Omega  {{\bf B}_\phi ^T{\bf g}{{\bf B}_\phi }d\Omega }, \\
  \\
  {{\bf{K}}_{u\phi }} = \int_\Omega  {{({\bf B}^L)}^T{{\bf \tilde{e}}^T}{{\bf B}_\phi }d\Omega },\;{{\bf{K}}_{\phi u}} = {\bf{K}}_{u\phi }^T,\;
  {\bf{M}_{uu}}{\rm{ = }}\int_\Omega  {{{{\bf{\tilde N}}}^T}{\bf m}{\bf{\tilde N}}d\Omega },\;{\bf{F}} = \int_\Omega  {{{ q}_0}{\bf{\bar N}}{\rm{d}}\Omega },
\end{array}
\label{eqn:General_FEM1}
\eeq
in which
\beq
\begin{array}{l}
  {\bf \tilde {e}} = [\begin{array}{*{20}{c}}
        {{\bf e}_m^T} & {z{\bf e}_m^T} & {f\left( z \right){\bf e}_m^T} & {{\bf e}_s^T} & {f'\left( z \right){\bf e}_s^T}
      \end{array}], \\ \\ {\bf{\tilde N}} = [\begin{array}{*{20}{c}}{{\bf N}^0}&{{\bf N}^1}&{{\bf N}^2}\end{array}]^T,\;\bar{\bf N} = \left[ {\begin{array}{*{20}{c}}
          0 & 0 & {{R_I}} & 0 & 0 & 0 & 0
        \end{array}} \right],
\end{array}
\eeq
where
\beq
\begin{array}{l}
  {\bf e}_m = \left[ {\begin{array}{*{20}{c}}
          0          & 0          & 0          \\
          0          & 0          & 0          \\
          {{e_{31}}} & {{e_{32}}} & {{e_{33}}}
        \end{array}} \right],{\bf e}_s = \left[ {\begin{array}{*{20}{c}}
          0          & {{e_{15}}} \\
          {{e_{15}}} & 0          \\
          0          & 0
        \end{array}} \right],
  {\bf N}^0 = \left[ {\begin{array}{*{20}{c}}
          {{R_I}} & 0       & 0       & 0 & 0 & 0 & 0 \\
          0       & {{R_I}} & 0       & 0 & 0 & 0 & 0 \\
          0       & 0       & {{R_I}} & 0 & 0 & 0 & 0
        \end{array}} \right], \\ \\{\bf N}^1 =  - \left[ {\begin{array}{*{20}{c}}
          0 & 0 & 0 & {{R_I}} & 0       & 0 & 0 \\
          0 & 0 & 0 & 0       & {{R_I}} & 0 & 0 \\
          0 & 0 & 0 & 0       & 0       & 0 & 0
        \end{array}} \right],\;{\rm{ }}
  {\bf N}^2 = \left[ {\begin{array}{*{20}{c}}
          0 & 0 & 0 & 0 & 0 & {{R_I}} & 0       \\
          0 & 0 & 0 & 0 & 0 & 0       & {{R_I}} \\
          0 & 0 & 0 & 0 & 0 & 0       & 0
        \end{array}} \right]
\end{array}
\eeq
and
\beq
{\bf m} = \left[ {\begin{array}{*{20}{c}}
        {{I_1}} & {{I_2}} & {{I_4}} \\
        {{I_2}} & {{I_3}} & {{I_5}} \\
        {{I_4}} & {{I_5}} & {{I_6}}
      \end{array}} \right]
\eeq
where ${I}_i$ with $i=1:6$ are defined as
\beq
\left( {I}_1,{I}_2,{ I}_3,{I}_4,{I}_5,{I}_6 \right) =  \int_{ - h/2}^{h/2}{\rho (z)\left( {1,z,{z^{\rm{2}}},f(z),zf(z),{f^2}(z)} \right)}{\rm{d}}z.
\eeq

It should be noted that the electric potential field $\bf E$ depends only on following to $z$ direction leads to the matrix ${\bf K}_{u\phi}$ in Eq. (\ref{eqn:General_FEM1}) can be rewritten as
\beq
{{\bf{K}}_{u\phi }} = \int_\Omega  {\left( {{{\left( {{\bf B}^m} \right)}^T}{\bf e}_m^T{\bf B}_\phi + z{{\left( {{\bf B}^{b1}} \right)}^T}{\bf e}_m^T{\bf B}_\phi + f(z){{\left( {{\bf B}^{b2}} \right)}^T}{\bf e}_m^T{\bf B}_\phi} \right)} d\Omega.
\eeq
Finally, by substituting the second equation into the first one of  Eq. (\ref{eqn:General_FEM}), yields
\beq
{{\bf{M}}_{uu}}{\bf{\ddot d}} + \left( {{{\bf{K}}_{uu}}+ {{\bf{K}}_{u\phi }}{\bf{K}}_{_{\phi \phi }}^{ - 1}{{\bf{K}}_{\phi u}}} \right){\bf{d}} = {\bf{F}} + {{\bf{K}}_{u\phi }}{\bf{K}}_{_{\phi \phi }}^{ - 1}{\bf{Q}}.
\label{eqn:Total_Eqn}
\eeq

\section{Active control analysis}
\label{sec:Active_Control}
\begin{figure}[h!]
  \centering
  \includegraphics[trim=2cm 7cm 2cm 6.8cm,clip=true,scale=0.75]{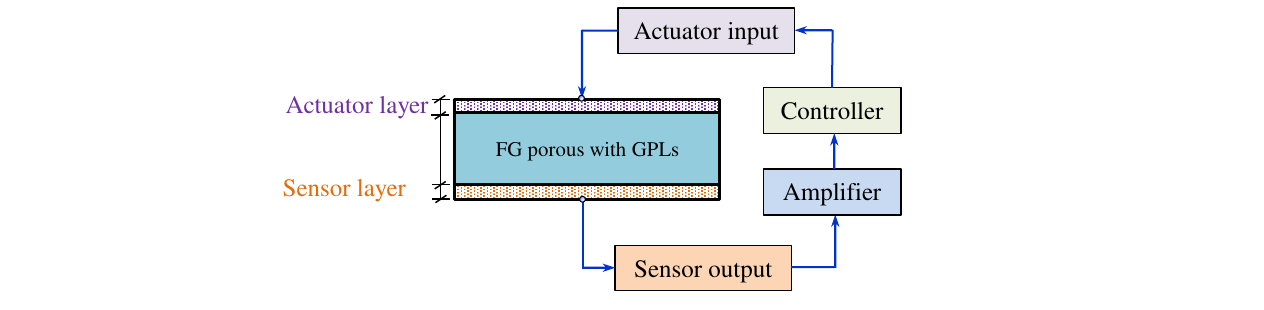}
  \caption{A schematic diagram of a FG porous with GPL reinforcement plate integrated two piezoelectric layers for active control process. The lower piezoelectric layer is referred to as the sensor layer while the other is considered as the actuator layer. A closed loop with added amplifier and controller components is established to connect the sensor and actuator layers.}
  \label{fig:Plate_Control}
\end{figure}

Consider a piezoelectric FG plate, as illustrated in Fig. \ref{fig:Plate_Control}, for the active control static and dynamic responses. While the lower layer is a piezoelectric sensor named with the subscript $s$, the upper one represents a piezoelectric actuator labeled with the subscript $a$. Because of the piezoelectric effect, the electric charge can be generated and collected in the sensor layer when the structures are deformed. After that, through an appropriate electronic circuit, this electric charge is amplified and converted into the voltage signal before being sent and applied to the actuator layer via closed loop control algorithm. Due to the converse piezoelectric phenomenon, the corresponding deformations can occur in structures which lead to damping the structural dynamic responses.\\

By ignoring the external charge $\bf Q$, the generated potential electric in sensor layer can be estimated based on the second equation of Eq. (\ref{eqn:General_FEM})
\beq
{\boldsymbol\phi} _s = {\left[ {{\bf{K}}_{\phi \phi }^{ - 1}} \right]_s}{\left[ {{\bf{K}}_{\phi u}^{}} \right]_s}{{\bf{d}}_s},
\label{eqn:Phi_s}
\eeq

Then, the sensor charge resulted because of the corresponding deformation can be expressed as
\beq
{{\bf{Q}}_s} = {\left[ {{\bf{K}}_{\phi u}} \right]_s}{{\bf{d}}_s}.
\eeq

In this study, two constant control gains of the displacement $G_d$ and velocity feedback $G_v$, respectively, are employed to couple the input and output voltage vectors as follows \cite{liu2004static}
\beq
{\boldsymbol\phi}_a=G_d{\boldsymbol\phi}_s+G_v{\dot{\boldsymbol\phi}}_s.
\label{eqn:Phi_a}
\eeq

By substituting Eqs. (\ref{eqn:Phi_a}) and (\ref{eqn:Phi_s}) into the second equation in Eq. (\ref{eqn:General_FEM}), the magnitude of actuator layer charge can be archived as follows
\beq
{{\bf{Q}}_a} = {\left[ {{\bf{K}}_{\phi u}^{}} \right]_a}{{\bf{d}}_a} - {G_d}{\left[ {{\bf{K}}_{\phi \phi }^{}} \right]_a}{\left[ {{\bf{K}}_{\phi \phi }^{ - 1}} \right]_s}{\left[ {{\bf{K}}_{\phi u}^{}} \right]_s}{{\bf{d}}_s} - {G_v}{\left[ {{\bf{K}}_{\phi \phi }^{}} \right]_a}{\left[ {{\bf{K}}_{\phi \phi }^{ - 1}} \right]_s}{\left[ {{\bf{K}}_{\phi u}^{}} \right]_s}{{\bf{\dot d}}_s}.
\label{eqn:Qa}
\eeq

Next, by substituting Eq. (\ref{eqn:Qa}) into Eq. (\ref{eqn:Total_Eqn}), one obtains
\beq
{\bf{M\ddot d}} + {\bf{C\dot d}} + {\bar{\bf{K}}}{\bf{d}} = {\bf{F}},
\label{eqn:Total_FEMwithC}
\eeq
in which
\beq
{\bar{\bf{K}}} = {\bf{K}}_{uu}^{} + {G_d}{\left[ {{\bf{K}}_{u\phi }^{}} \right]_a}{\left[ {{\bf{K}}_{\phi \phi }^{ - 1}} \right]_s}{\left[ {{\bf{K}}_{\phi u}^{}} \right]_s},
\eeq
and the active damping matrix $\bf C$ is expressed as
\beq
{\bf{C}} = {G_v}{\left[ {{\bf{K}}_{u\phi }} \right]_a}{\left[ {{\bf{K}}_{\phi \phi }^{ - 1}} \right]_s}{\left[ {{\bf{K}}_{\phi u}^{}} \right]_s}.
\eeq

Considering the structural damping effect, Eq. (\ref{eqn:Total_FEMwithC}) can be rewritten as
\beq
{\bf{M\ddot d}} + {({\bf C}+{\bf C}_R)\dot {\bf d}} + {\bar{\bf{K}}}{\bf{d}} = {\bf{F}},
\eeq
where the Rayleigh damping matrix ${\bf C}_R$ is determined based on a linear combination between $\bf M$ and ${\bf K}_{uu}$, as follows
\beq
{\bf C}_R=\alpha_R{\bf M}+\beta_R{\bf K}_{uu},
\eeq
in which $\alpha_R$ and $\beta_R$ denote Rayleigh damping coefficients which can obtain from experiments.

\section{Numerical examples}
\label{sec:num_exam}
\begin{figure}[h!]
  \centering
  \includegraphics[trim=1.0cm 7cm 1.8cm 6.5cm,clip=true,scale=0.45]{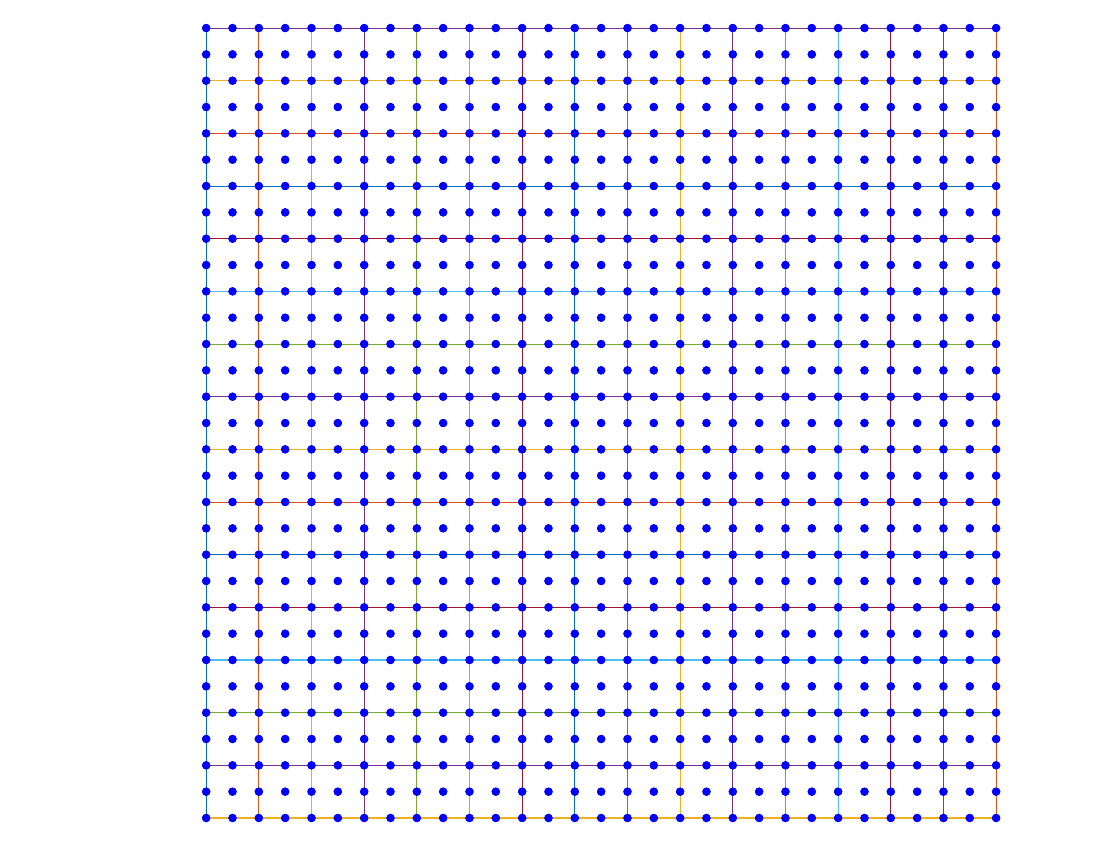}
  \caption{A B\'ezier control mesh ($15 \times 15$) for a FG square plate.}
  \label{fig:Mesh_15x15}
\end{figure}

In this work, Newton-Rapshon iterative procedure is employed to obtain the solution of nonlinear problems. For the geometrically nonlinear dynamic analysis of FG plate under various dynamic loads, which the equations depend on both the time domain and variable displacement, Newmark's integration scheme is exploited. In all numerical examples, the PZT-G1195N piezoelectric is employed as well as perfectly glued on the upper and lower surfaces of plate and neglected the adhesive layers. In addition, a mesh of $15\times15$ quadratic B\'ezier elements, as illustrated in Fig. \ref{fig:Mesh_15x15}, is utilized in order to model all square plate structures in this work.

\subsection{Validation analysis}
In this section, various numerical studies regarding the geometrically nonlinear static and dynamic analyses of isotropic as well as piezoelectric FG plates are carried out in order to perform the accuracy of the proposed approach. For geometrically nonlinear analysis, an isotropic square plate with a fully clamped (CCCC) boundary conditions is considered to show the validity of the present formulation. The plate structure is subjected to a uniform transverse load while the width-to-thickness ratio $(a/h)$ is taken equal to $100$. The material properties are $E=3\times10^7$ psi and $\nu=0.316$. In this example, the normalized central deflection and load parameter are given as $\overline{w}=w/h$ and $P=q_0a^4/(Eh^2)$, respectively. Table \ref{tab:ISO_CCCC} presents the normalized central deflections which are compared with those of Levy's analytical solution \cite{levy1942square}, Urthaler and Reddy's mixed FEM using FSDT \cite{urthaler2008mixed} and Nguyen et al. based on IGA and refined plate theory \cite{nguyen2017geometrically}. As can be observed that the present results are in good agreement with the existing analytical solution as well as other approximate results.\\
\begin{table}[h!]
  \centering
  \caption{Normalized central deflection $\overline{w}$ of a CCCC isotropic square plate ($a/h=100$) under a uniform load ($P=q_0a^4/(Eh^2)$).}
  \scalebox{0.9}{\begin{tabular}{ccccclllll}
      \hline
      $P$   & Present & Analytical \cite{levy1942square} & MXFEM \cite{urthaler2008mixed} & IGA-RPT \cite{nguyen2017geometrically} \\
      \hline
      17.79 & 0.2348  & 0.237                            & 0.2328                         & 0.2365                                 \\
      38.3  & 0.4663  & 0.471                            & 0.4738                         & 0.4692                                 \\
      63.4  & 0.6873  & 0.695                            & 0.6965                         & 0.6908                                 \\
      95.0  & 0.8983  & 0.912                            & 0.9087                         & 0.9024                                 \\
      134.9 & 1.1016  & 1.121                            & 1.1130                         & 1.1060                                 \\
      184.0 & 1.2960  & 1.323                            & 1.3080                         & 1.3008                                 \\
      245.0 & 1.4875  & 1.521                            & 1.5010                         & 1.4926                                 \\
      318.0 & 1.6728  & 1.714                            & 1.6880                         & 1.6784                                 \\
      402.0 & 1.8492  & 1.902                            & 1.8660                         & 1.8552                                 \\
      \hline
      \label{tab:ISO_CCCC}
    \end{tabular}}
\end{table}
\begin{figure}[h!]
  \centering
  \includegraphics[trim=1.4cm 6.5cm 1.8cm 7.2cm,clip=true,scale=0.47]{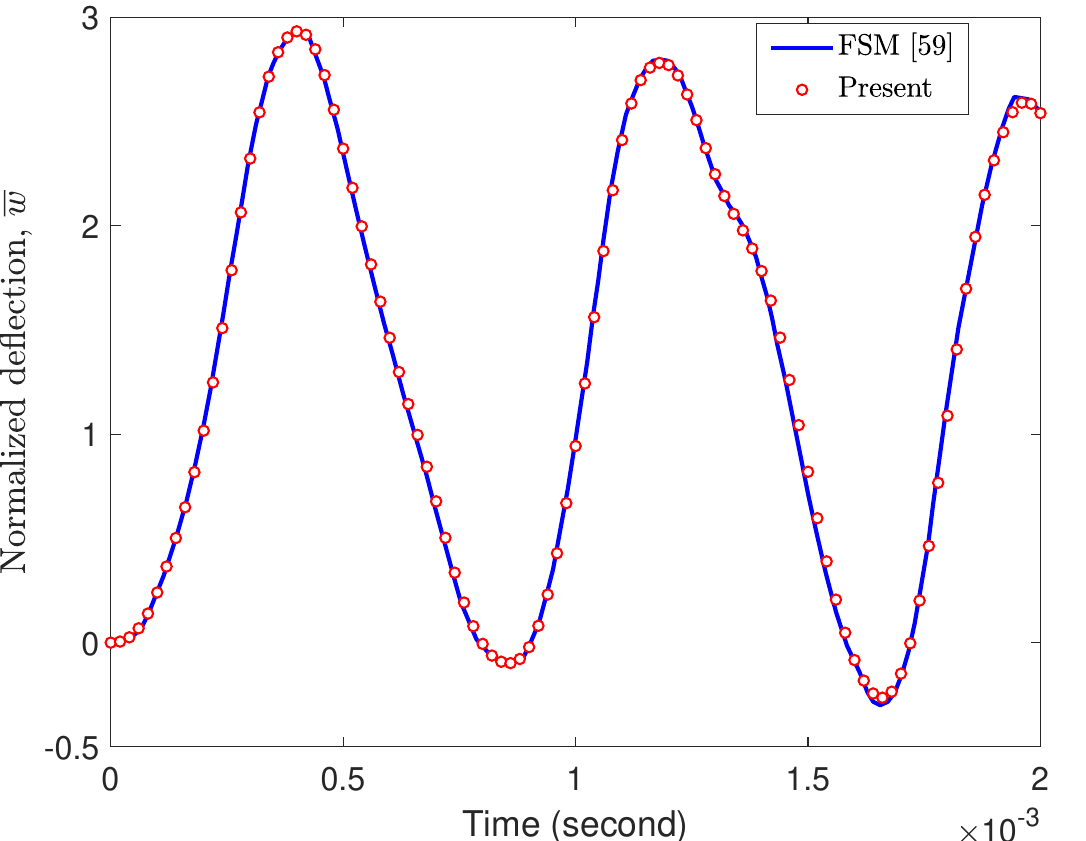}
  \caption{Geometrically nonlinear dynamic response of a SSSS orthotropic square plate ($a = b =250$ mm and $h = 5$ mm) under a uniform load $q_0 = 1.0$ MPa.}
  \label{fig:Nonlinear_Iso}
\end{figure}

Next, in order to verify the accuracy of the proposed approach for the geometrically nonlinear dynamic analysis, a fully simply supported (SSSS) boundary conditions orthotropic square plate subjected to a uniform load with $q_0=1.0$ MPa is conducted. The material properties and the geometry dimensions are given as follows: Young's modulus $E_1=525$ GPa, $E_2=21$ GPa, shear modulus $G_{12}=G_{23}= G_{13}=10.5$ GPa, Poisson's ratio $\nu=0.25$, mass density $\rho=800$ kg/m$^3$, length $a = b =250$ mm and thickness $ h=5$ mm. Fig. \ref{fig:Nonlinear_Iso} depicts the geometrically nonlinear dynamic response of square plate. It can be observed that the present result matchs very well with that obtained from the finite strip method, as reported by Chen at al. \cite{chen2000nonlinear}.\\
\begin{table}[htbp]
  \centering
  \caption{Material properties of the metal and piezoelectric materials.}
  \scalebox{0.75}{\begin{tabular}{llllllllll}
      \hline
      Properties                 & Core layer &                &        &        &   & {\textrm {Piezoelectric layer }}         \\
      \cline{2-5} \cline{7-7}
                                 & Ti-6Al-4V  & Aluminum oxide & Copper & GPLs   &   & PZT-G1195N                               \\
      \cline{2-7}
      \hline
      Elastic properties         &            &                &        &        &                                              \\
      $E_{11}$ (GPa)             & 105.70     & 320.24         & 130    & 1010   &   & 63.0                                     \\
      $E_{22}$ (GPa)             & 105.70     & 320.24         & 130    & 1010   &   & 63.0                                     \\
      $E_{33}$ (GPa)             & 105.70     & 320.24         & 130    & 1010   &   & 63.0                                     \\
      $G_{12}$ (GPa)             & $-$        & $-$            & $-$    & $-$    &   & 24.2                                     \\
      $G_{13}$ (GPa)             & $-$        & $-$            & $-$    & $-$    &   & 24.2                                     \\
      $G_{23}$ (GPa)             & $-$        & $-$            & $-$    & $-$    &   & 24.2                                     \\
      $\nu_{12}$                 & 0.2981     & 0.26           & 0.34   & 0.186  &   & 0.30                                     \\
      $\nu_{13}$                 & 0.2981     & 0.26           & 0.34   & 0.186  &   & 0.30                                     \\
      $\nu_{23}$                 & 0.2981     & 0.26           & 0.34   & 0.186  &   & 0.30                                     \\
      Mass density               &            &                &        &        &   &                                  &  &  & \\
      $\rho$  (kg/m$^3$)         & 4429       & 3750           & 8960   & 1062.5 &   & 7600                                     \\
      Piezoelectric coefficients &            &                &        &        &   &                                          \\
      $k_{31}$ (m/V)             & $-$        & $-$            & $-$    & $-$    &   & $254 \times {10^{ - 12}}$                \\
      $k_{32}$ (m/V)             & $-$        & $-$            & $-$    & $-$    &   & $254 \times {10^{ - 12}}$                \\
      Electric permittivity      &            &                &        &        &                                              \\
      $p_{11}$ (F/m)             & $-$        & $-$            & $-$    & $-$    &   & $15.3 \times {10^{ - 9}}$                \\
      $p_{22}$ (F/m)             & $-$        & $-$            & $-$    & $-$    &   & $15.3 \times {10^{ - 9}}$                \\
      $p_{33}$ (F/m)             & $-$        & $-$            & $-$    & $-$    &   & $15.3 \times {10^{ - 9}}$                \\
      \hline
    \end{tabular}}
  \label{tab:Table_Material}
\end{table}

\begin{figure}[h!]
  \centering
  \begin{subfigure}[c]{0.475\textwidth}
    \centering
    \includegraphics[trim=1.2cm 6.2cm 1.8cm 6cm,clip=true,scale=0.45]{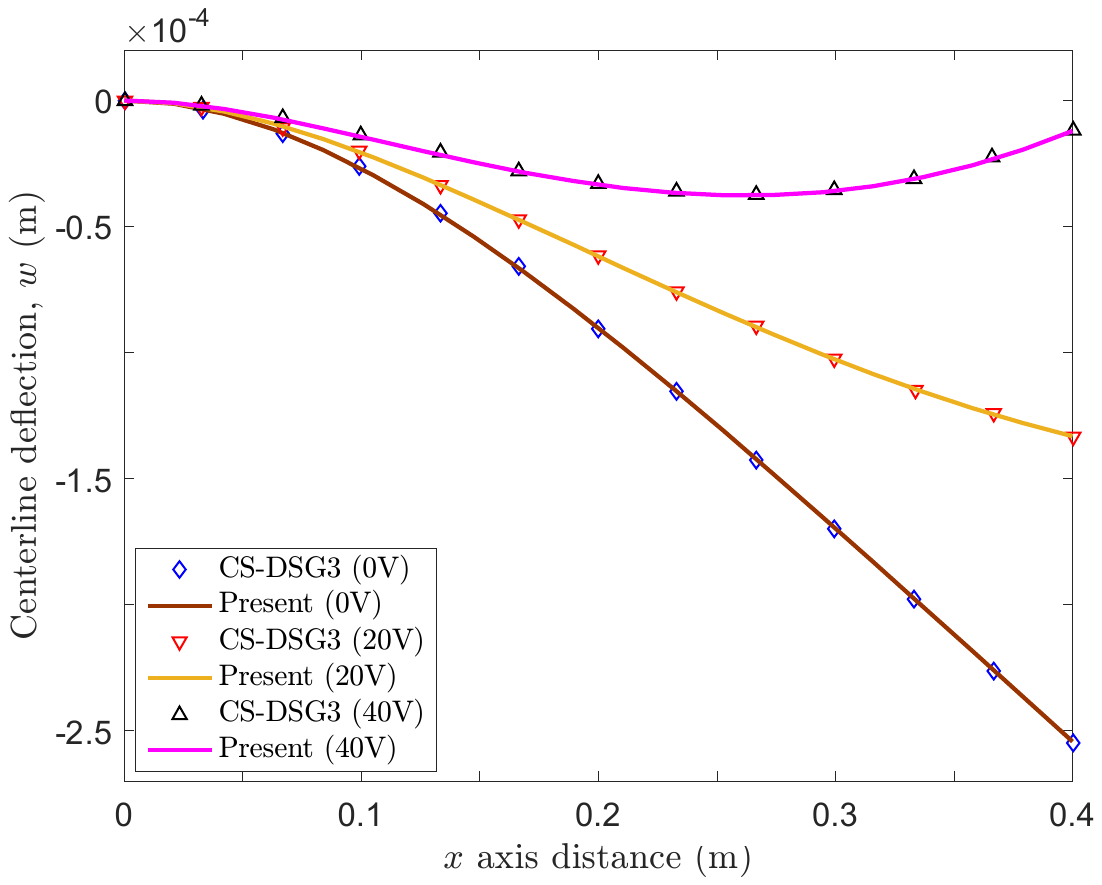}
    \caption{}%
  \end{subfigure}
  \hfill
  \begin{subfigure}[c]{0.475\textwidth}
    \centering
    \includegraphics[trim=1.2cm 6.2cm 1.8cm 6cm,clip=true,scale=0.45]{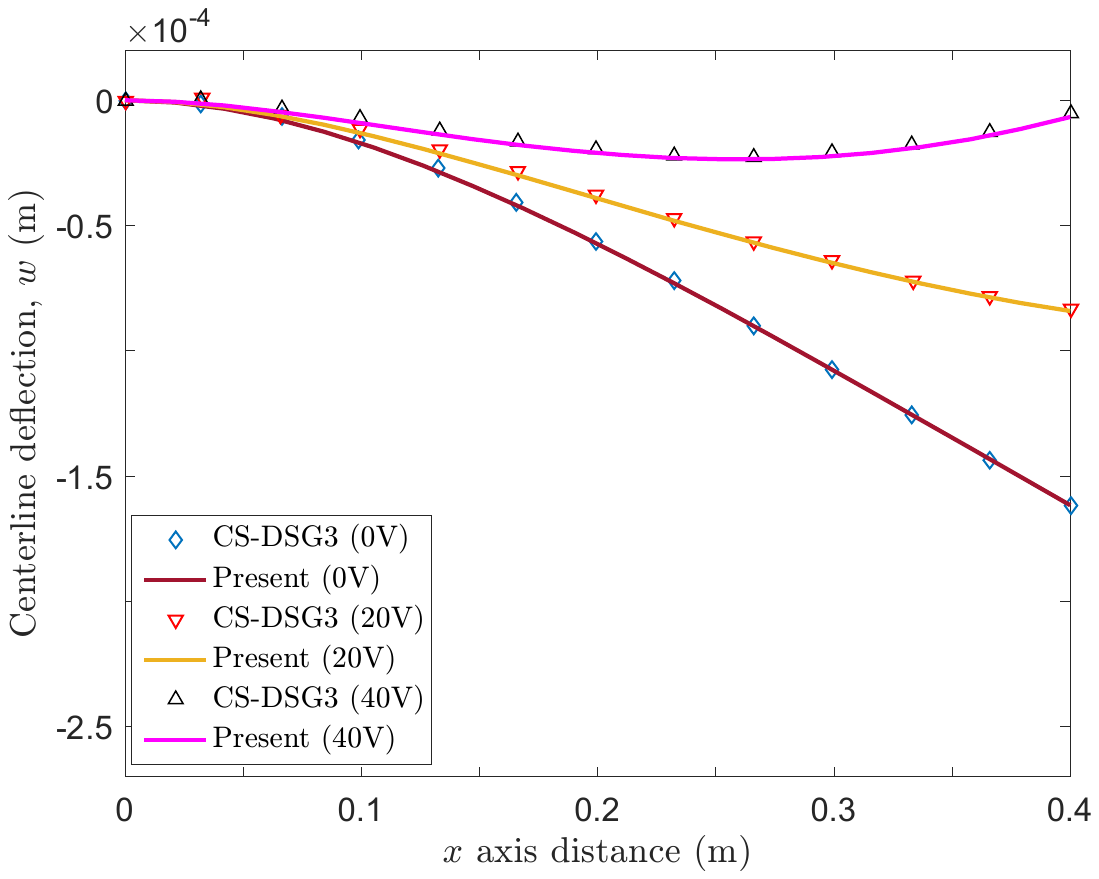}
    \caption{}%
  \end{subfigure}
  \hfill
  \caption{Centerline linear deflection of a cantilever FGM (Aluminum oxide/Ti-6Al-4V) plate ($a = b = 0.4$ m, $h_c = 5$ mm and $h_p = 0.1$ mm) integrated two piezoelectric layers under simultaneously a uniform mechanical load ($q_0 = 100$ N/m$^2$) and various input voltages (electrical load): (a) $n=0.0$; (b) $n=0.5$. The reference solution is reported in \cite{nguyen2017analysis} using CS-DSG3 based on FSDT.}
  \label{fig:Compare_CSDSG3}
\end{figure}

Regarding the plate structures integrated with piezoelectric layers, a cantilever piezoelectric FG square plate is thoroughly considered to verify the accuracy of the present approach. The FG plate which is bonded by two piezoelectric layers on both the lower and upper surfaces is made of aluminum oxide and Ti-6Al-4V materials and is subjected to simultaneously a uniform transverse load with $q_0$=100 N/m$^2$ and various input voltage values. The mechanical properties of two materials are given in Table \ref{tab:Table_Material}. In this study, the rule of mixture \cite{nakamura2000determination} is employed to calculate the effective material properties of FG core layer. The FG plate has a side length $a=b=0.4$ m whilst the thickness of core layer and each piezoelectric layer are $h_c=5$ mm and  $h_p=0.1$ mm, respectively. The centerline linear deflections of piezoelectric FG square plate are plotted in Fig. \ref{fig:Compare_CSDSG3} while the tip node deflections are also listed in Table \ref{tab:Deflec_Tip} with various material index $n$. The obtained results are compared with those reported in \cite{nguyen2017analysis} based on a cell-based smoothed discrete shear gap method (CS-DSG3) using FSDT. It can be observed that the present results generally agree well with the reference solutions. \\
In the next Sections, investigations into the geometrically nonlinear static and dynamic responses of piezoelectric FG porous plate with GPLs reinforcement will be presented.

\begin{table}[h!]
  \centering
  \caption{Tip node deflection of cantilever piezoelectric FGM (Aluminum oxide/Ti-6Al-4V) plate ($a = b = 0.4$ m, $h_c = 5$ mm and $h_p = 0.1$ mm) under a uniform load ($q_0 = 100$ N/m$^2$) and various input voltages ($\times 10^{-4}$ m).}
  \scalebox{0.9}{\begin{tabular}{llllllllllll}
      \hline
      $n$        & Method                            & \multicolumn{3}{l}{\textrm {Input voltages (V)}} &                   \\
      \cline{3-6}
                 &                                   & 0                                                & 20      & 40      \\
      \hline
      $n=0$      & Present                           & -2.5437                                          & -1.3328 & -0.1229 \\
                 & CS-DSG3 \cite{nguyen2017analysis} & -2.5460                                          & -1.3346 & -0.1232 \\
      $n=0.5$    & Present                           & -1.6169                                          & -0.8418 & -0.0667 \\
                 & CS-DSG3 \cite{nguyen2017analysis} & -1.6199                                          & -0.8440 & -0.0681 \\
      $n=5$      & Present                           & -1.1233                                          & -0.5808 & -0.0382 \\
                 & CS-DSG3 \cite{nguyen2017analysis} & -1.1266                                          & -0.5820 & -0.0375 \\
      $n=\infty$ & Present                           & -0.8946                                          & -0.4608 & -0.0271 \\
                 & CS-DSG3 \cite{nguyen2017analysis} & -0.8947                                          & -0.4609 & -0.0271 \\
      \hline
      \label{tab:Deflec_Tip}
    \end{tabular}}
\end{table}
\subsection{Geometrically nonlinear static analysis}
\label{sec:Geo_Static}
\begin{figure}[h!]
  \centering
  \includegraphics[trim=0cm 6.5cm 0cm 7cm,clip=true,scale=0.47]{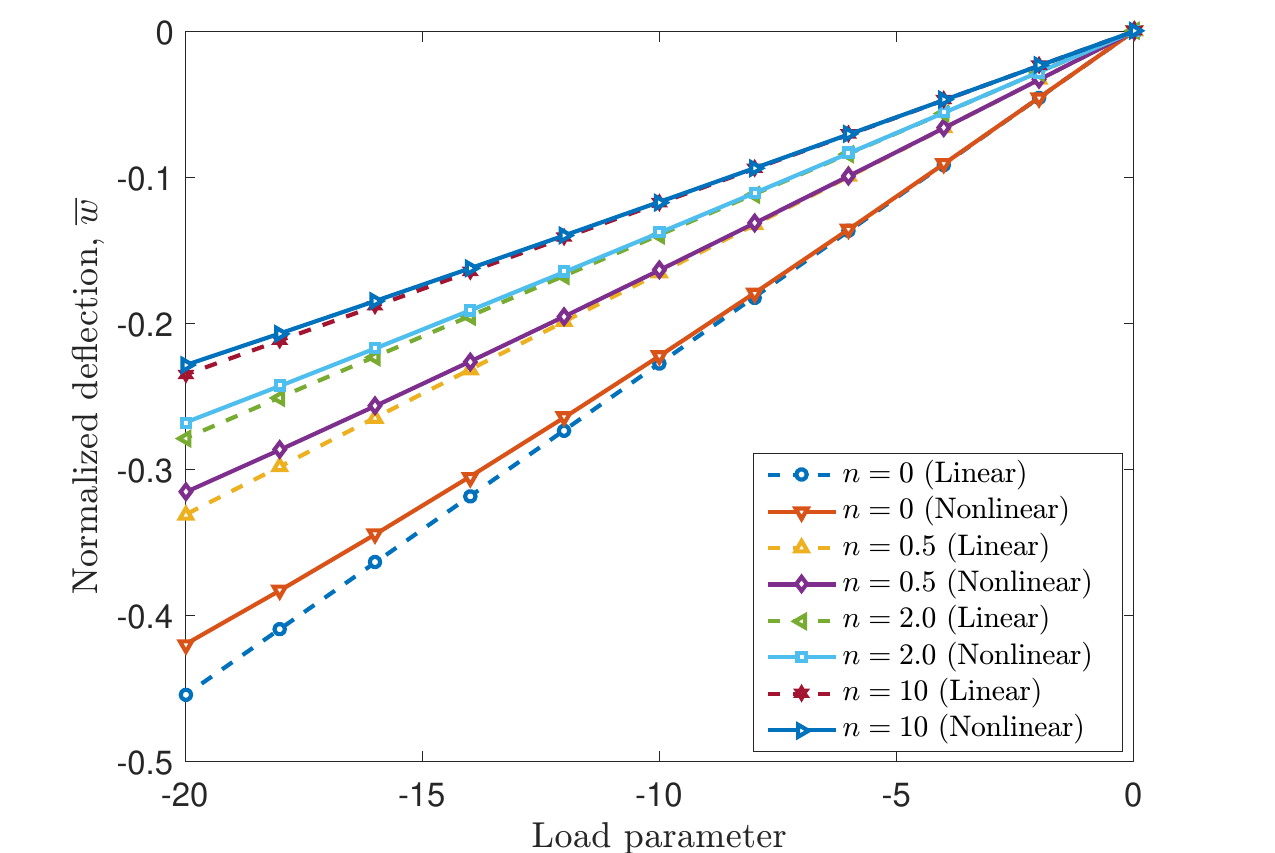}
  \caption{Influence of material index $n$ on linear and geometrically nonlinear responses of a SSSS FGM (Aluminum oxide/Ti-6Al-4V) plate ($a = b = 0.2$ m, $h_c = 2.0$ mm and $h_p = 0.1$ mm) integrated two piezoelectric layers under only uniform mechanical load. Note that $n = 0.0$ leads to a fully homogeneous ceramic material (Ti-6Al-4V).}
  \label{fig:Nonlinear_Deflec}
\end{figure}

The geometrically nonlinear static analysis of a piezoelectric FG plate subjected to uniform load with $\overline{q} = q_0\times 10^3$ is first addressed. A SSSS piezoelectric FG square plate made of aluminum oxide and Ti-6Al-4V has $a =b= 0.2$ m, $h_c= 2.0$ mm and $h_p = 0.1$ mm. Fig. \ref{fig:Nonlinear_Deflec} illustrates the influence of material index $n$ on the normalized linear and nonlinear central deflections of piezoelectric FG plates under mechanical load. As can be observed that with an increase of material index $n$ leads to a decrease gradually of the deflection. The largest deflection occurs with $n=0.0$ where the plate includes only of Ti-6Al-4V leads to a decrease in the bending stiffness. Furthermore, the values of the central deflection for geometrically nonlinear analysis are always smaller than that of the linear one. This difference would gradually reduce with the increase of material index.\\

\begin{figure}[h!]
  \centering
  \begin{subfigure}[c]{0.475\textwidth}
    \centering
    \includegraphics[trim=1.4cm 6.2cm 1.8cm 6.7cm,clip=true,scale=0.47]{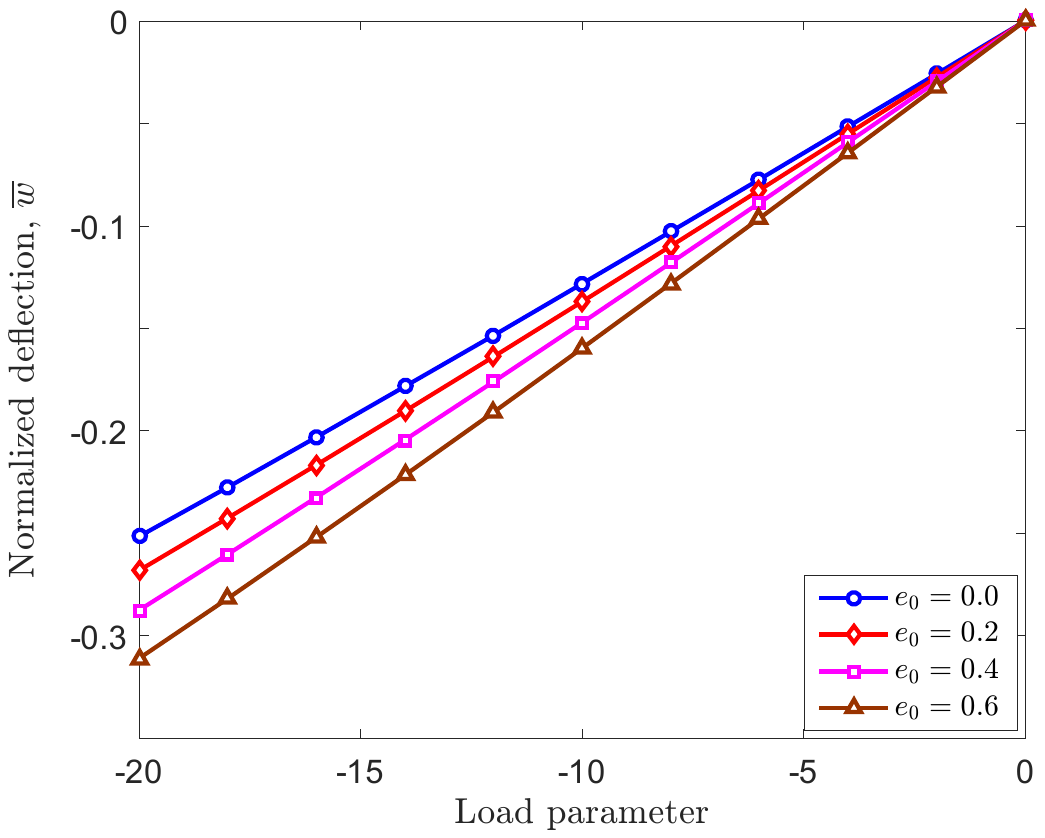}
    \caption{}%
  \end{subfigure}
  \hfill
  \begin{subfigure}[c]{0.475\textwidth}
    \centering
    \includegraphics[trim=1.4cm 6.2cm 1.8cm 6.7cm,clip=true,scale=0.47]{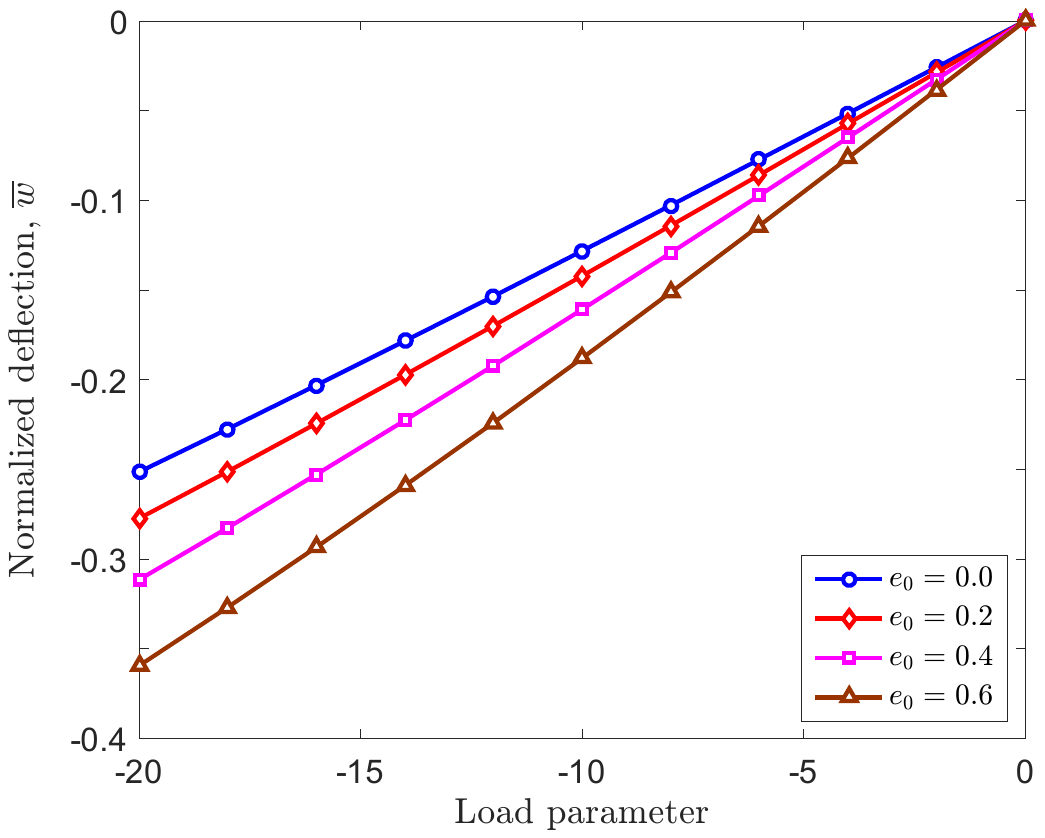}
    \caption{}%
  \end{subfigure}
  \hfill
  \caption{Influence of porosity coefficient $e_0$ on nonlinear response of a SSSS piezoelectric FG porous plate ($a = b = 0.4$ m, $h_c = 20$ mm and $h_p = 1.0$ mm) whose the core layer is made of two porosity distributions and GPL pattern A ($\Lambda_{GPL}=1.0 \; wt.\;\%$), respectively: (a) porosity distribution 1; (b) porosity distribution 2. Note that, $e_0 = 0.0$ implies no porosities in metal matrix. }
  \label{fig:Compare_P12}
\end{figure}

In the next example, we consider a SSSS piezoelectric FG plate having core layer constituted by a combination of two porosity distribution types and three GPL patterns, respectively. The FG plate is subjected to sinusoidally distributed load defined as $q=q_0sin(\pi x/a)sin(\pi y/b)$ in which $q_0=1.0$ MPa. The geometry dimensions of plate are taken as $a =b= 0.4$ m, $h_c= 20$ mm and $h_p = 1.0$ mm. In this study, the copper is chosen as the metal matrix whose material properties are given in Table \ref{tab:Table_Material}. Meanwhile, the basic dimensions of GPLs are ${l_{GPL}} = 2.5\;\mu m$, ${w_{GPL}} = 1.5\;\mu m$, ${t_{GPL}} = 1.5\;nm$. Fig. \ref{fig:Compare_P12} examines the influence of porosity coefficient on the nonlinear normalized deflection of piezoelectric FG porous plate with pattern $A$ $(\Lambda_{GPL}=1.0\; wt.\;\%)$ for two porosity distributions, respectively. It can be observed that an increase of the porosity coefficient leads to the corresponding increase of the nonlinear deflection since the higher density of internal pores in material yields the reduction in the stiffness of structures.

\begin{figure}[h!]
  \centering
  \begin{subfigure}[c]{0.475\textwidth}
    \centering
    \includegraphics[trim=1.4cm 6.2cm 1.8cm 7.4cm,clip=true,scale=0.47]{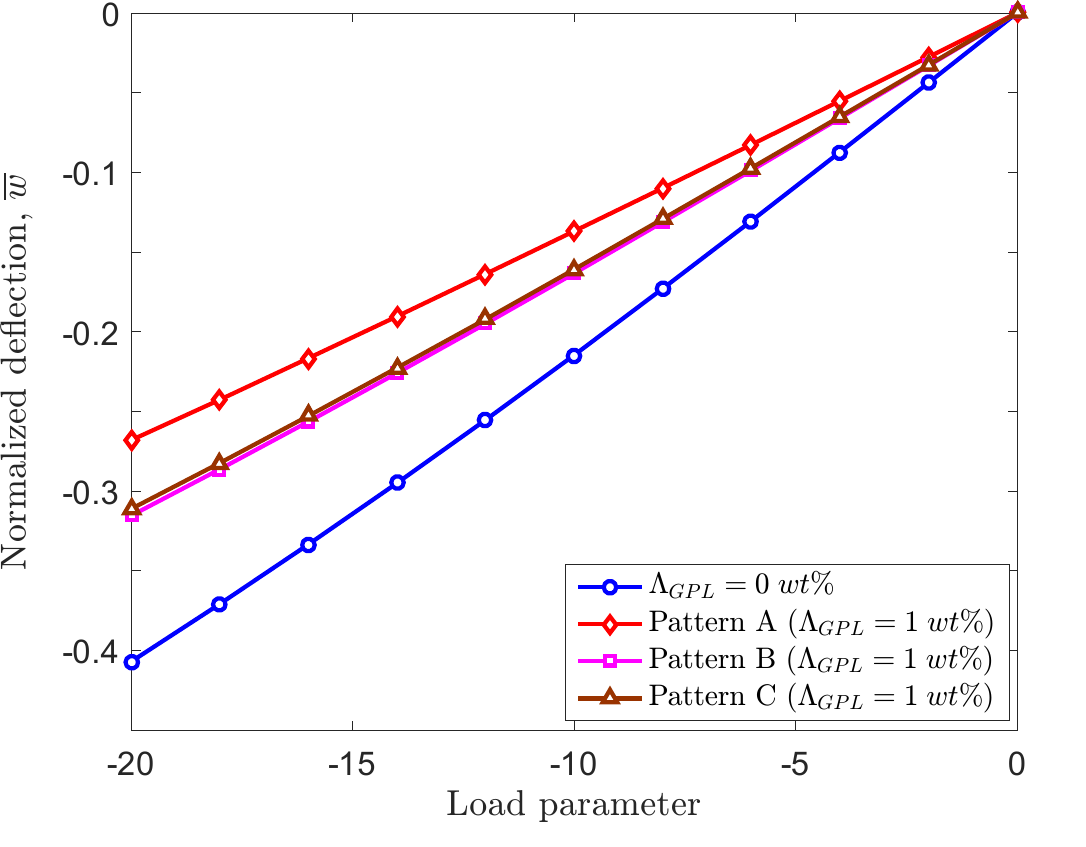}
    \caption{}
  \end{subfigure}
  \hfill
  \begin{subfigure}[c]{0.475\textwidth}
    \centering
    \includegraphics[trim=1.4cm 6.2cm 1.8cm 7.4cm,clip=true,scale=0.47]{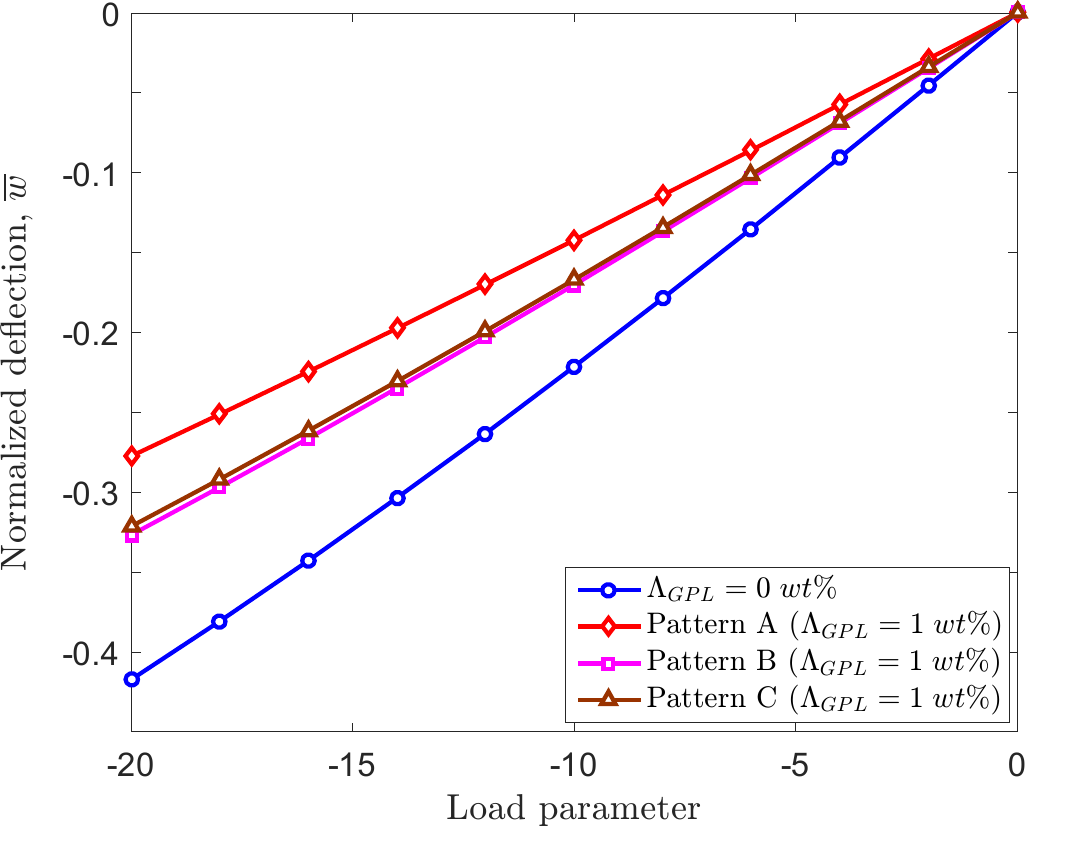}
    \caption{}
  \end{subfigure}
  \hfill
  \caption{Influence of weight fraction $\Lambda_{GPL}$ and patterns of GPLs on nonlinear response of a SSSS piezoelectric FG porous plate ($a = b = 0.4$ m, $h_c = 20$ mm and $h_p = 1.0$ mm) where the core layer is constituted by two porosity distributions with $e_0=0.2$: (a) porosity distribution 1; (b) porosity distribution 2. $\Lambda_{GPL} = 0.0$ implies no GPL reinforcement in metal matrix. }
  \label{fig:Compare_PatternGPL}
\end{figure}

\begin{figure}[H]
  \centering
  \begin{subfigure}[b]{0.475\textwidth}
    \centering
    \includegraphics[trim=0.8cm 7cm 0.9cm 7.4cm,clip=true,scale=0.4]{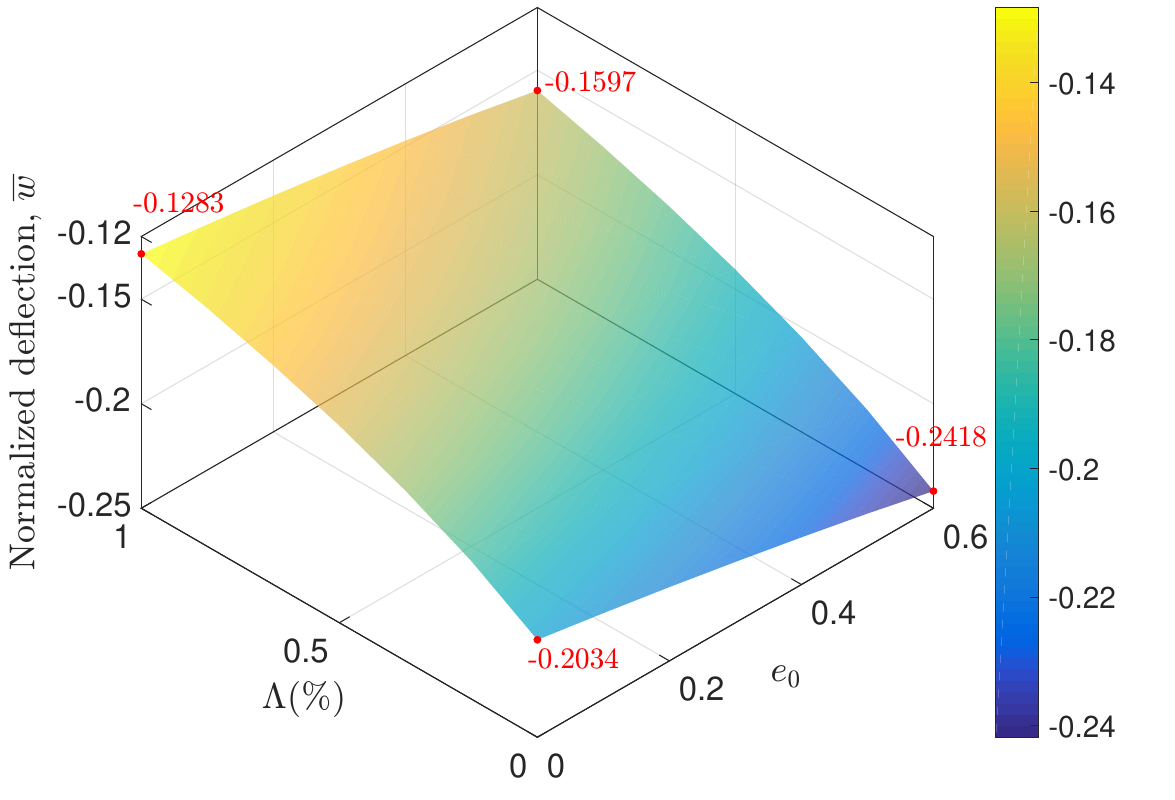}
    \caption{}%
  \end{subfigure}
  \begin{subfigure}[b]{0.475\textwidth}
    \centering
    \includegraphics[trim=0.7cm 7cm 1.0cm 7.4cm,clip=true,scale=0.4]{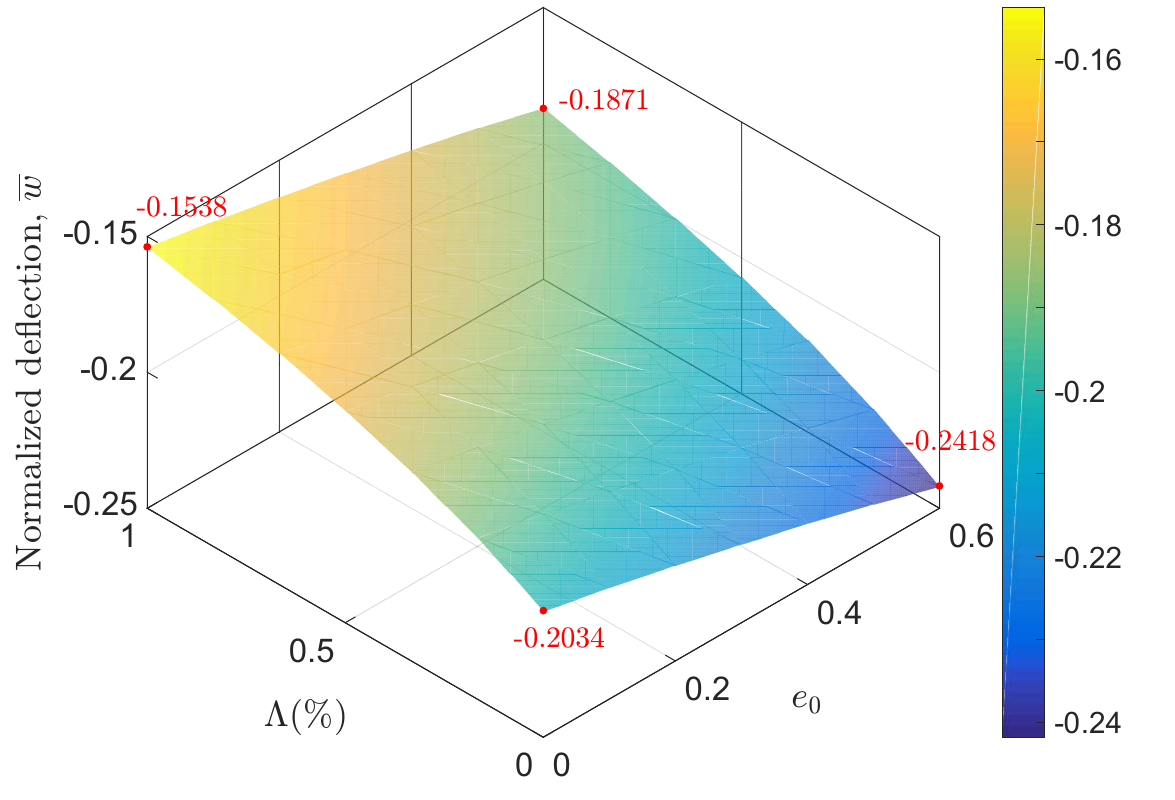}
    \caption{}%
  \end{subfigure}
  \begin{subfigure}[b]{0.475\textwidth}
    \centering
    \includegraphics[trim=0.7cm 7cm 1cm 6.5cm,clip=true,scale=0.4]{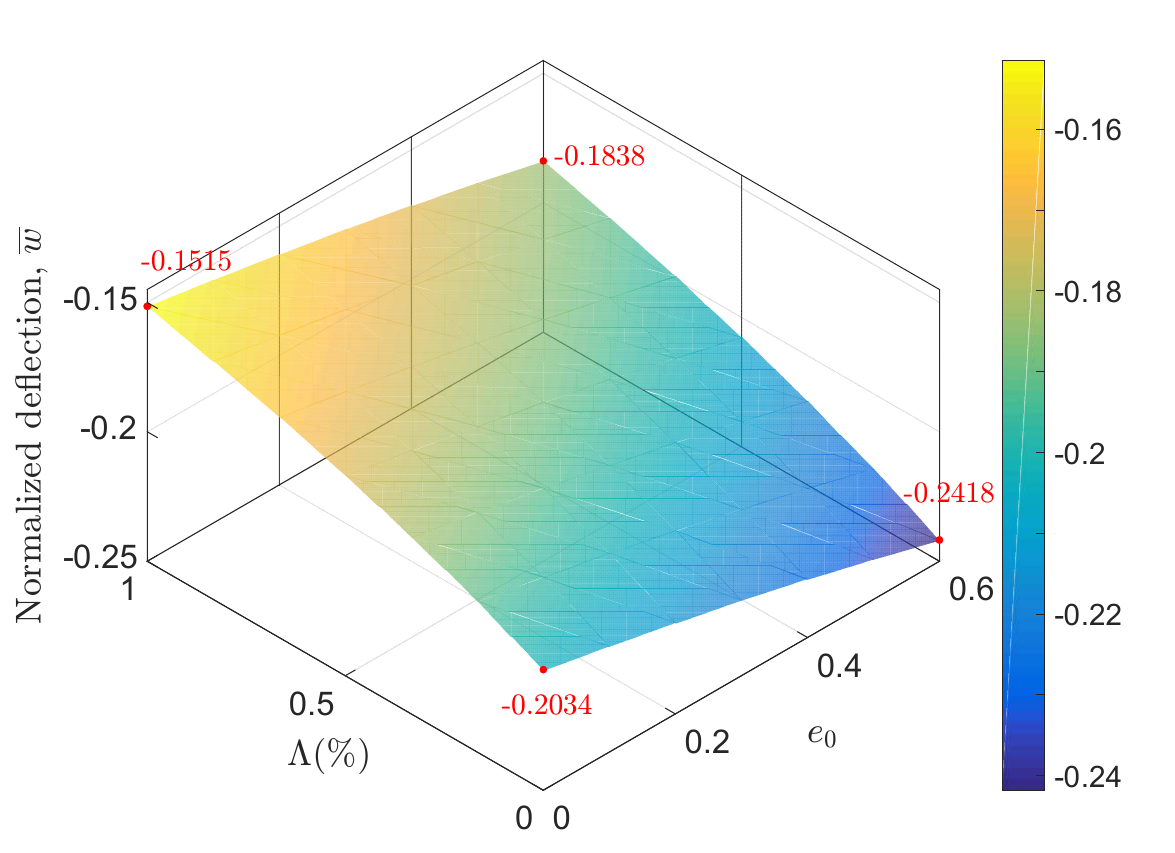}
    \caption{}%
  \end{subfigure}
  \caption{Influence of porosity coefficient $e_0$ and weight fraction of GPLs $\Lambda_{GPL}$ on nonlinear responses of a SSSS piezoelectric FG porous plate ($a = b = 0.4$ m, $h_c = 20$ mm and $h_p = 1.0$ mm). The core layer is constituted by a combination between porosity distribution 1 and three GPL patterns: (a) pattern A; (b) pattern B; (c) pattern C. The results are generated corresponding to the load parameter $P = 10$.}
  \label{fig:Non_P1_G123}
\end{figure}

In addition, Fig. \ref{fig:Compare_PatternGPL} depicts the influence of weight fraction and GPL dispersion patterns on the behaviors of FG porous plate with $e_0=0.2$ and two porosity distributions, respectively. We observe that the effective stiffness of FG porous core layer is significantly strengthened after adding a small amount of GPLs ($\Lambda_{GPL}=1.0\;wt.\;\% )$ into metal matrix as evidenced by a decrease of the nonlinear normalized deflection. More importantly, the reinforcement effect of GPLs depends significantly on the dispersion type of GPLs into metal matrix. Accordingly, with the same weight fraction of GPLs, the pattern $A$, where GPLs are dispersed symmetrically through the midplane of porous core layer, achieves the smallest nonlinear deflection while the asymmetric pattern $B$ provides the largest one. For further illustration, Fig. \ref{fig:Non_P1_G123} depicts the variation of nonlinear normalized deflection of piezoelectric FG porous plate with GPLs reinforcement which is constituted by a combination of porosity distribution 1 and three different GPL dispersion patterns, respectively with load parameter 10.\\

\begin{figure}[H]
  \centering
  \includegraphics[trim=0cm 6.5cm 0cm 7.5cm,clip=true,scale=0.47]{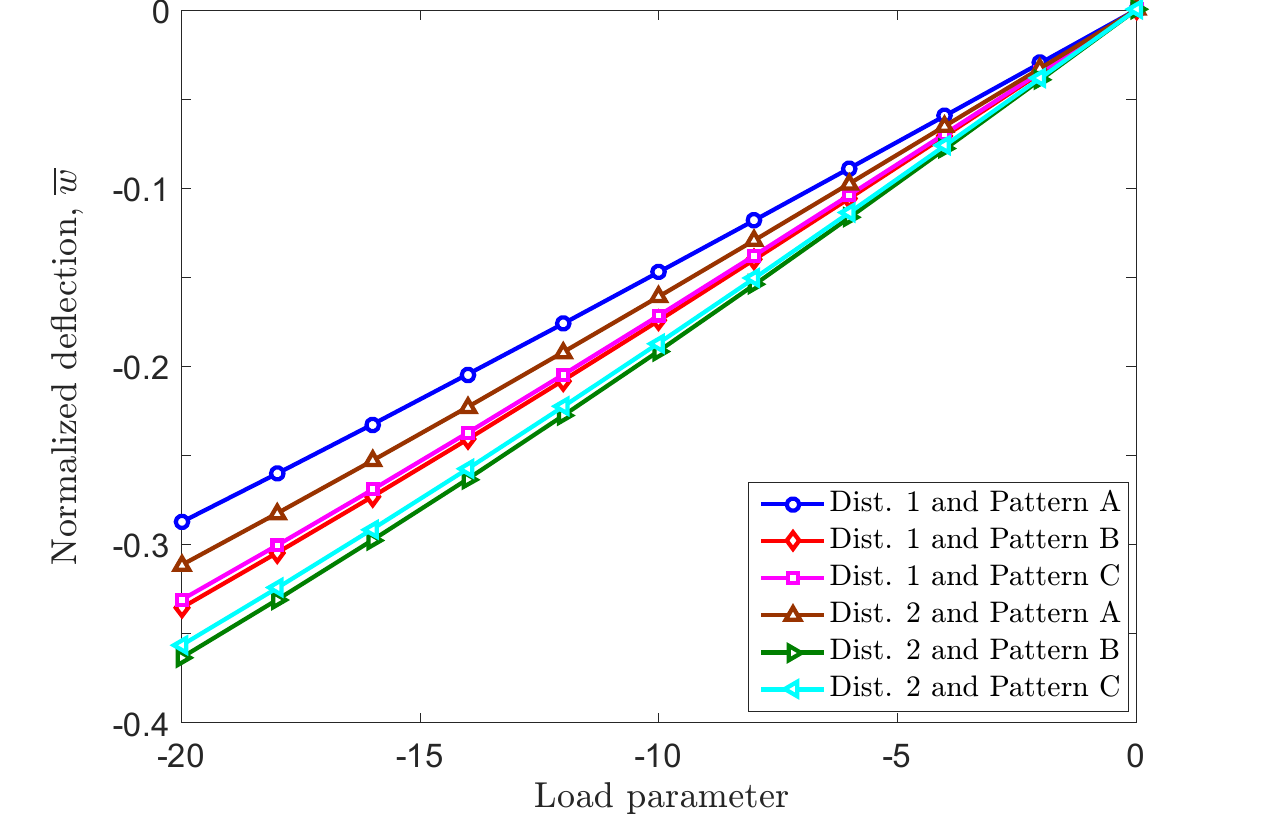}
  \caption{Influence of distribution of porosity with $e_0=0.4$ and GPLs with $\Lambda_{GPL}=1.0 \;wt.\;\%$ on nonlinear response of a SSSS piezoelectric FG porous plate ($a = b = 0.4$ m, $h_c = 20$ mm and $h_p = 1.0$ mm).}
  \label{fig:Nonlinear_P_G}
\end{figure}

The combined influences of two porosity distributions and three GPL patterns on the nonlinear deflection of FG plate structure with $\Lambda_{GPL}=1.0\; wt.\;\%$ and $e_0=0.4$ is thoroughly examined. As evidently depicted in Fig. \ref{fig:Nonlinear_P_G}, for all the considered cases, the combination of porosity distribution 1 and pattern $A$ always obtains the best reinforcement performance in the geometrically nonlinear static problems. This indicated that the FG plate structures where the internal pores are dispersed on the midplane and GPLs are distributed around the bottom and top surfaces yield the optimum reinforcement. \\

\subsection{Geometrically nonlinear dynamic analysis}
\label{sec:Transient}
\begin{figure}[h!]
  \centering
  \includegraphics[trim=0cm 7.8cm 0cm 8.2cm,clip=true,scale=0.55]{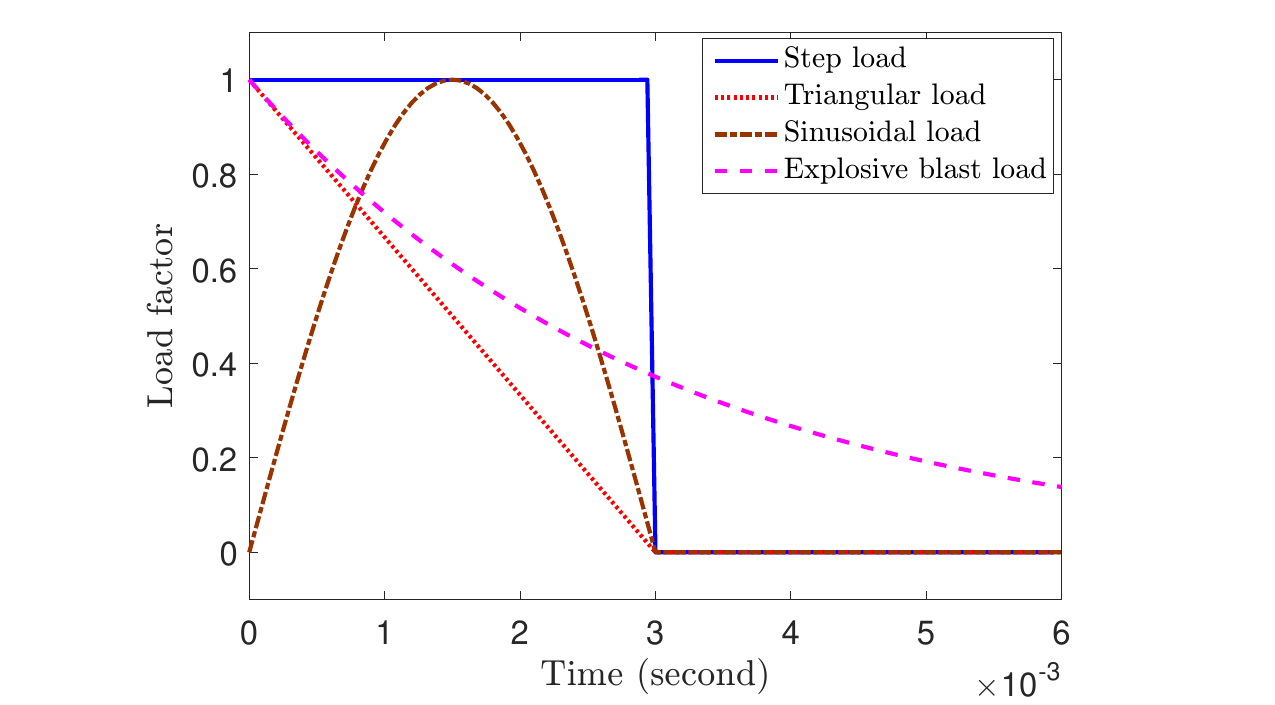}
  \caption{Time history curves of load factor $F(t)$ for four dynamic loads including step, triangular, sinusoidal and explosive blast loads, as defined in Eq. (\ref{eqn:load}). $t_1 = 3 \times 10^{-3}$ second is the time which the step, triangular and sinusoidal loads are suddenly removed.}
  \label{fig:Loading}
\end{figure}

In this part, the geometrically nonlinear dynamic behaviors of a CCCC piezoelectric FG porous plate with GPL reinforcement are investigated. The basic dimensions as well as the material properties of FG plate are the same previous example. The FG plate is assumed to be subjected to time-dependent sinusoidally distributed transverse loads expressed as follows $q=q_0sin(\pi x/a)sin(\pi y/b)F(t)$. In this work, four different dynamic loads are examined via the function $F(t)$ as follows
\beq
{F}\left( t \right) = \left\{ {\begin{array}{*{20}{c}}
      {\left\{ {\begin{array}{*{20}{c}}
              1 \\
              0
            \end{array}\,\,\,\,\,\,\,\,\,\,\,\,\,\,\,\,\,\,\,} \right.} & {\begin{array}{*{20}{c}}
            {0 \le t \le {t_1},} \\
            {\,\,\,\,\,\,\,t > {t_1},}
          \end{array}} & {{\rm{Step\; load}}\,\,\,\,\,\,\,\,\,\,\,\,\,\,} \\
      {\left\{ {\begin{array}{*{20}{c}}
            {1 - t/{t_1}\,\,\,\,\,\,} \\
            {0\,\,\,\,\,\,\,\,\,\,\,\,\,\,\,\,\,\,\,}
          \end{array}} \right.}                                       & {\begin{array}{*{20}{c}}
            {0 \le t \le {t_1},} \\
            {\,\,\,\,\,\,\,t > {t_1},}
          \end{array}} & {{\rm{Triangular \;load}}\,\,\,}                 \\
      {\left\{ {\begin{array}{*{20}{c}}
            {{\rm{sin}}\left( {\pi t/{t_1}} \right)} \\
            0
          \end{array}} \right.}                                       & {\begin{array}{*{20}{c}}
            {0 \le t \le {t_1},} \\
            {\,\,\,\,\,\,\,t > {t_1},}
          \end{array}} & {{\rm{Sinusoidal\; load}}\,\,\,}                 \\
      {{e^{ - \gamma t}},\,\,\,\,\,\,\,\,\,\,\,\,\,\,\,\,}                                 & {}                           & {\,\,\,\,\,\,\,{\rm{Explosive\; blast\; load}}}
    \end{array}} \right.
    \label{eqn:load}
\eeq
in which $q_0=100$ MPa, $\gamma=330s^{-1}$. Fig. \ref{fig:Loading} illustrates the time history $F(t)$ with $t_1 = 3\times10^{-3}$ s.\\

\begin{figure}[h!]
  \centering
  \begin{subfigure}[c]{0.49\textwidth}
    \centering
    \includegraphics[trim=1.2cm 6.8cm 0.5cm 7.2cm,clip=true,scale=0.470]{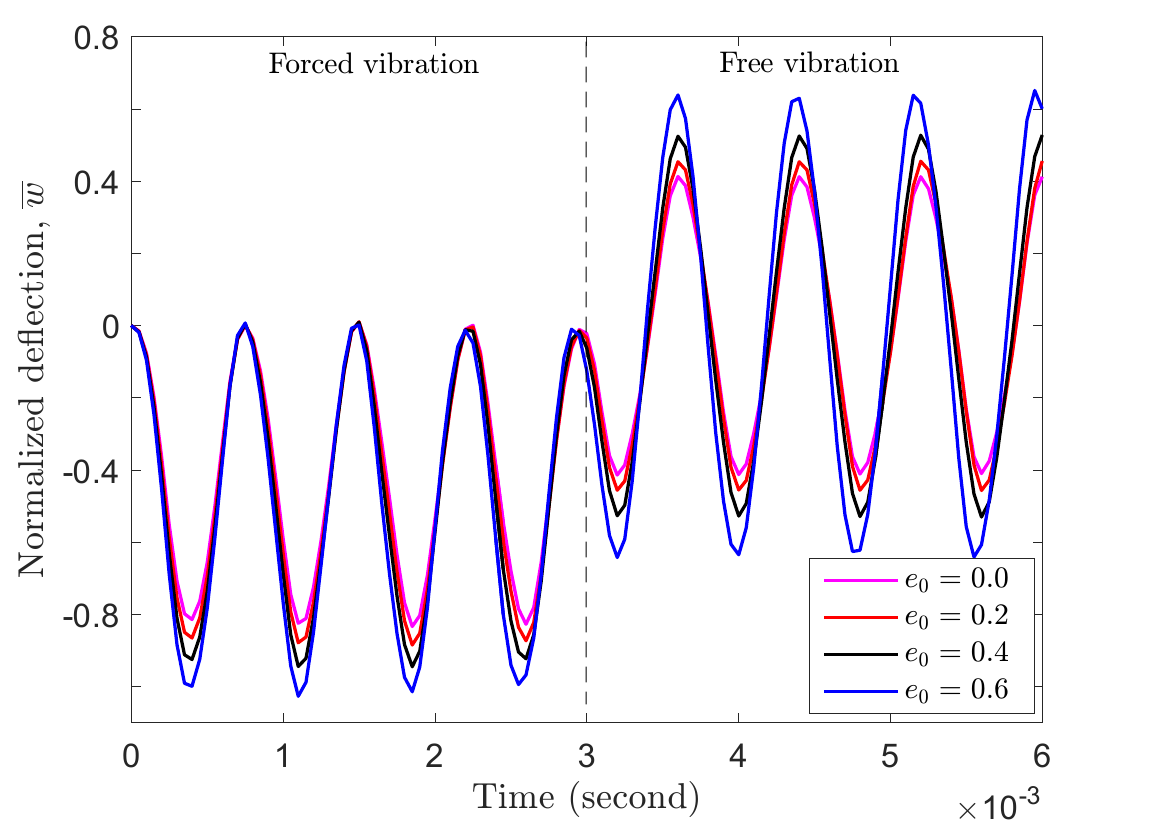}
    \caption{}%
  \end{subfigure}
  \hfill
  \centering
  \begin{subfigure}[c]{0.49\textwidth}
    \centering
    \includegraphics[trim=1.2cm 6.8cm 0.5cm 7.2cm,clip=true,scale=0.470]{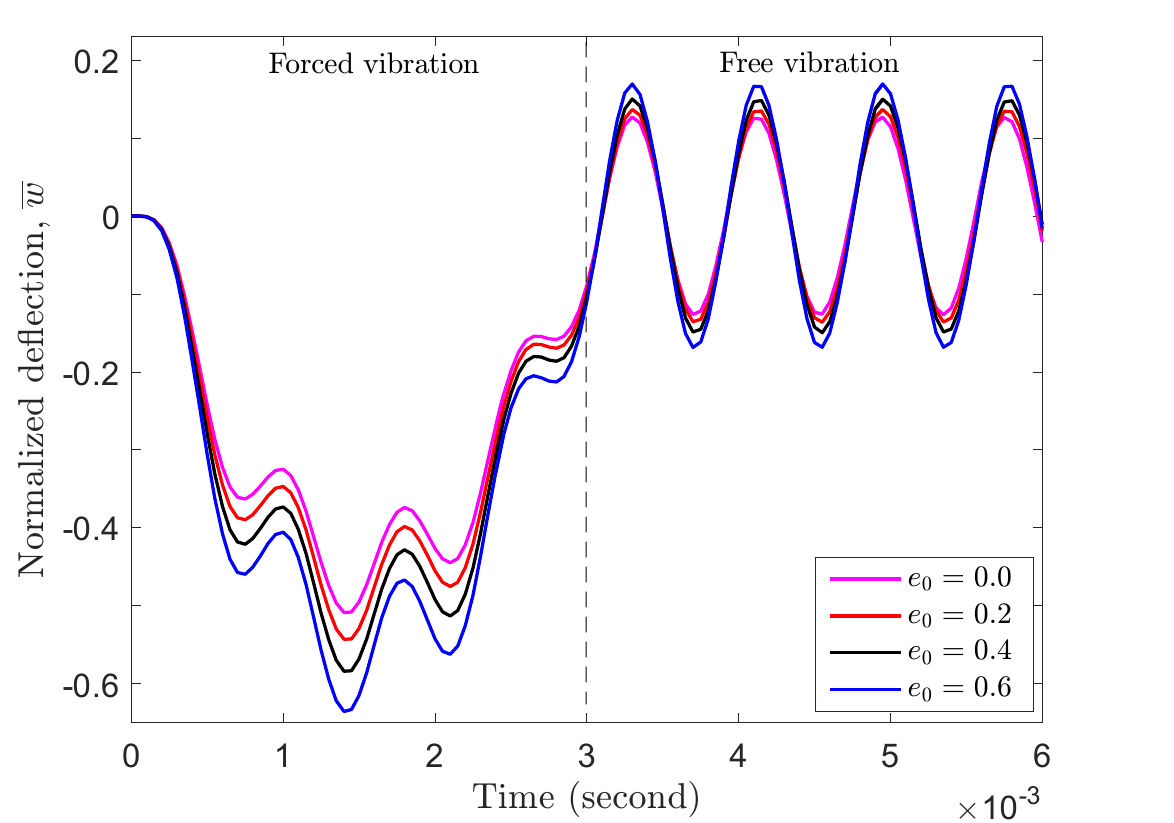}
    \caption{}%
  \end{subfigure}
  \caption{Influence of porosity coefficient $e_0$ on nonlinear dynamic responses of a CCCC piezoelectric FG porous plate ($a = b = 0.4$ m, $h_c = 20$ mm and $h_p = 1.0$ mm) having porosity distribution A and GPL pattern A ($\Lambda_{GPL}=1.0\; wt.\;\%$): (a) step load; (b) sinusoidal load. The FG plate is subjected to dynamic loads, as defined in Eq. (\ref{eqn:load}), in the time interval from 0 to $t_1 = 3 \times 10^{-3}$ second and then is free vibration. Note that, $e_0 = 0.0$ refer to no pores in metal matrix.}
  \label{fig:Transient_e}
\end{figure}

\begin{figure}[h!]
  \centering
  \begin{subfigure}[c]{0.49\textwidth}
    \centering
    \includegraphics[trim=1.2cm 6.8cm 0.5cm 7.2cm,clip=true,scale=0.470]{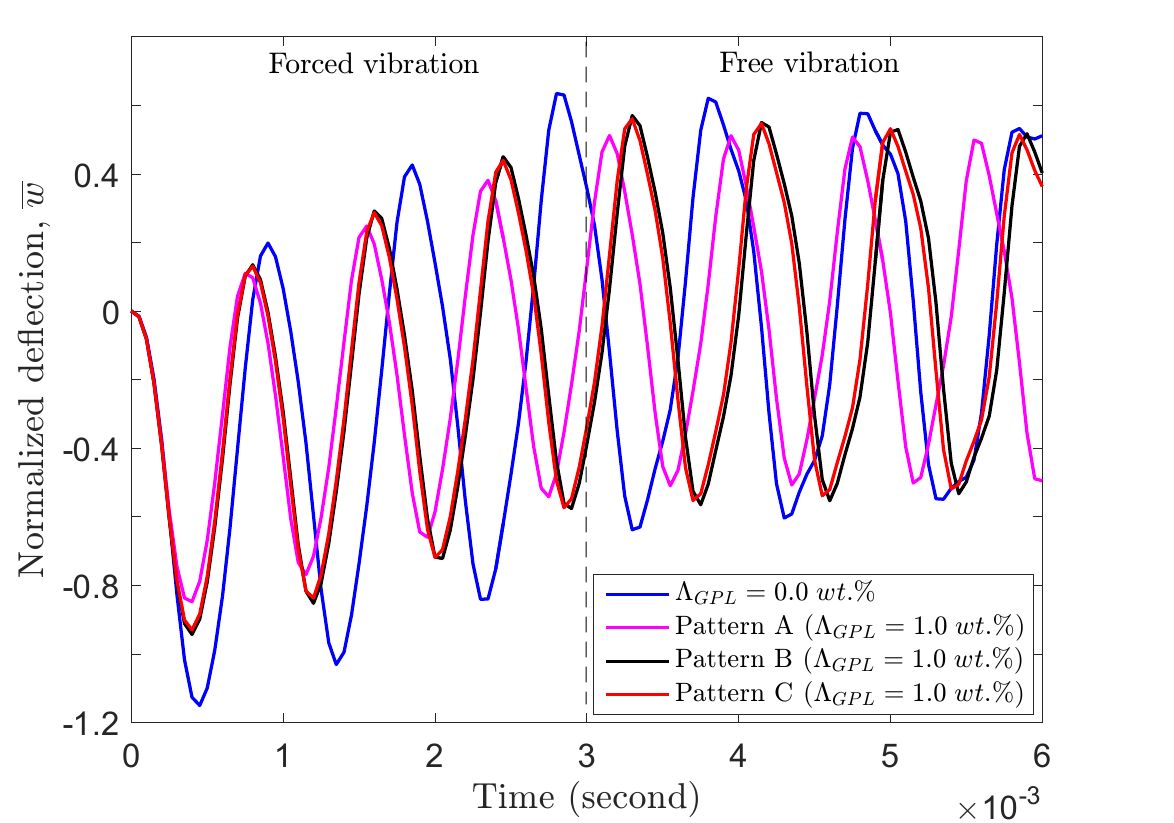}
    \caption{}%
  \end{subfigure}
  \hfill
  \centering
  \begin{subfigure}[c]{0.49\textwidth}
    \centering
    \includegraphics[trim=1.2cm 6.8cm 0.5cm 7.2cm,clip=true,scale=0.470]{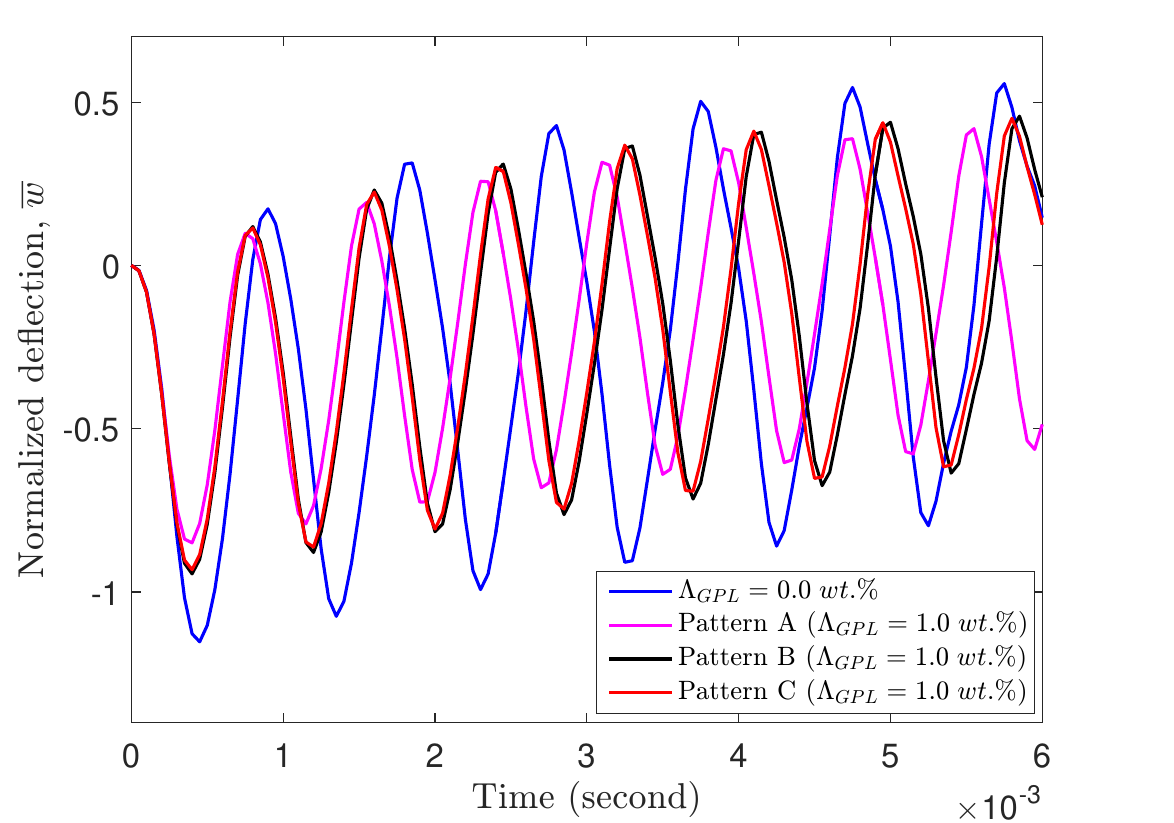}
    \caption{}%
  \end{subfigure}
  \caption{Influence of weight fraction $\Lambda_{GPL}$ and pattern of GPLs on nonlinear dynamic responses of CCCC piezoelectric FG porous plate ($a = b = 0.4$ m, $h_c = 20$ mm and $h_p = 1.0$ mm) with porosity distribution 2 ($e_0=0.2$): (a) triangular load; (b) explosive blast load. The FG plate is subjected to triangular load, as defined in Eq. (\ref{eqn:load}), in the time interval from 0 to $t_1 = 3 \times 10^{-3}$ second and then is free vibration. $\Lambda_{GPL} = 0.0$ implies no GPL reinforcement in metal matrix.}
  \label{fig:Transient_GPL}
\end{figure}

Fig. \ref{fig:Transient_e} illustrates the influence of porosity coefficient on the nonlinear dynamic response of  piezoelectric FG porous plate with porosity distribution 1 and pattern $A$ $(\Lambda_{GPL}=1.0\; wt.\;\%)$ under step and sinusoidal loads. It can be observed that when the porosity coefficient increases, the amplitude of the transverse deflection increases as well while the period of motion does not seem to affect. This is because the presence of porosities in core layer leads to reduction the capacity of itself against external excitation. Furthermore, Fig. \ref{fig:Transient_GPL} performs the influence of weight fraction and GPL dispersion pattern on the nonlinear dynamic response of piezoelectric FG porous plate with $e_0=0.2$ and porosity distribution 2 corresponding to triangular and explosive blast loads. As expected, the smaller magnitude of the deflection can obtain when the weight fraction of GPLs into metal matrix increases. Again, the dispersion of GPLs into metal matrix significantly influence to the reinforcement performance of structure that dispersion pattern $A$ provides the smallest magnitude of the deflection.\\

\begin{figure}[h!]
  \centering
  \begin{subfigure}[c]{0.49\textwidth}
    \centering
    \includegraphics[trim=4.9cm 2.7cm 3.5cm 1.2cm,clip=true,scale=0.410]{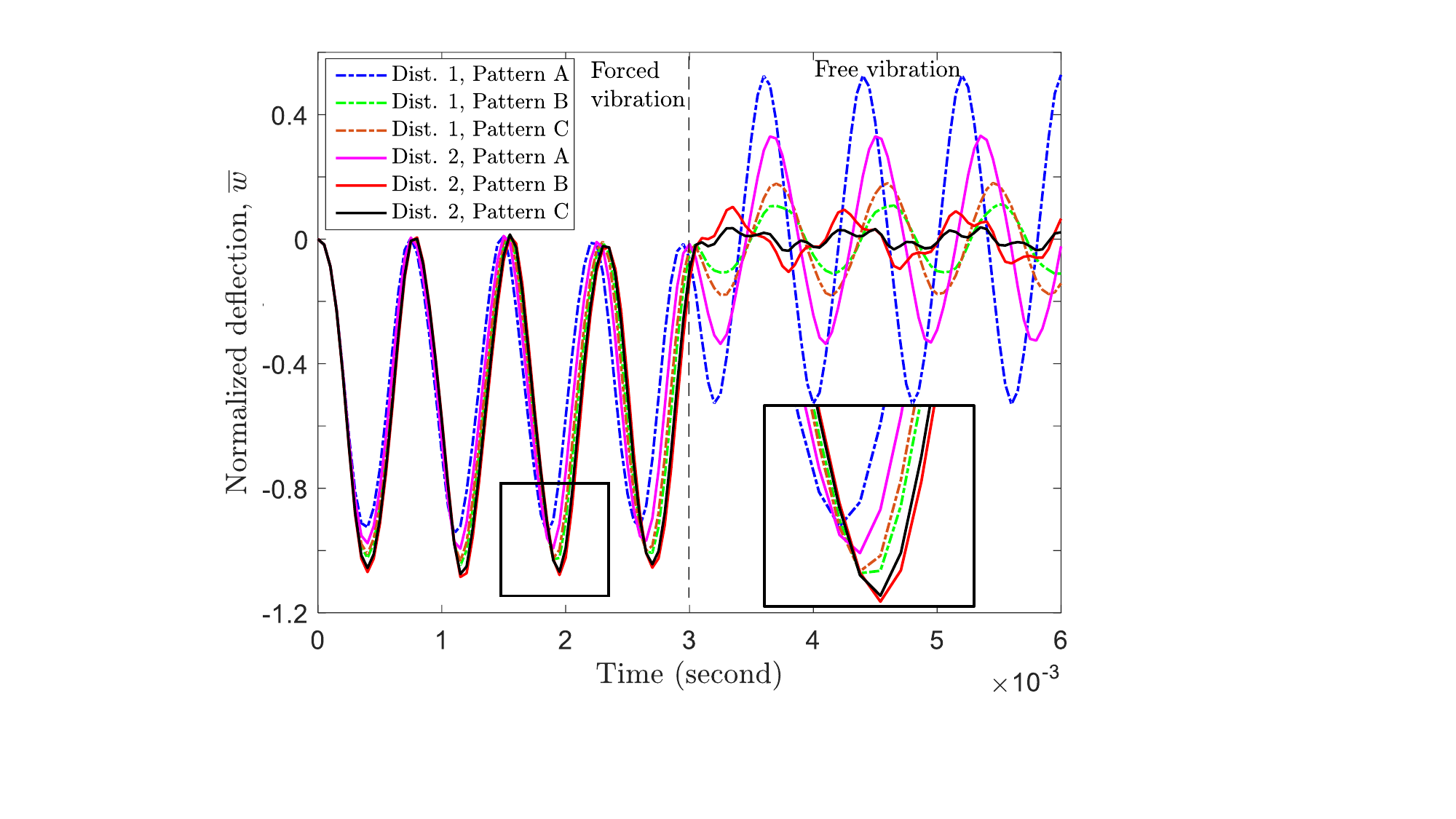}
    \caption{}%
  \end{subfigure}
  \hfill
  \centering
  \begin{subfigure}[c]{0.49\textwidth}
    \centering
    \includegraphics[trim=4.7cm 0.6cm 4cm 2.0cm,clip=true,scale=0.380]{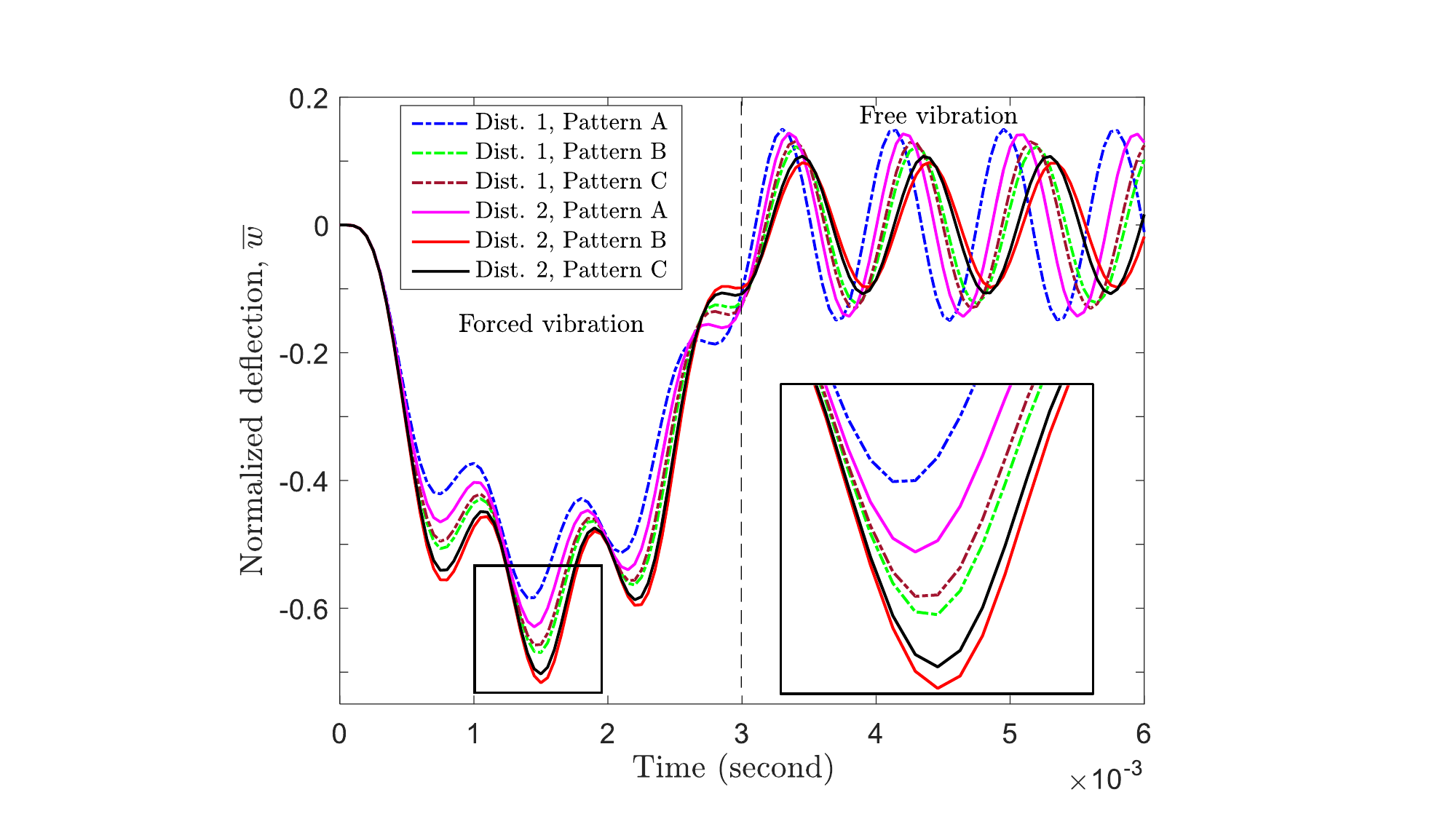}
    \caption{}%
  \end{subfigure}
  \caption{Influence of distribution of porosity and GPL on nonlinear dynamic responses of CCCC piezoelectric FG porous plate ($a = b = 0.4$ m, $h_c = 20$ mm and $h_p = 1.0$ mm) with $e_0=0.4$ and $\Lambda_{GPL}=1.0\; wt.\;\%$: (a) step load; (b) sinusoidal load. The FG plate is subjected to dynamic loads, as defined in Eq. (\ref{eqn:load}), in the time interval from 0 to $t_1 = 3 \times 10^{-3}$ second and then is free vibration. The zoom-in of the selected peak is presented in lower right corner of each figure.}
  \label{fig:Transient_P_GPL}
\end{figure}

\begin{figure}[h!]
  \centering
  \begin{subfigure}[c]{0.475\textwidth}
    \centering
    \includegraphics[trim=1.4cm 6.5cm 1.8cm 6.5cm,clip=true,scale=0.47]{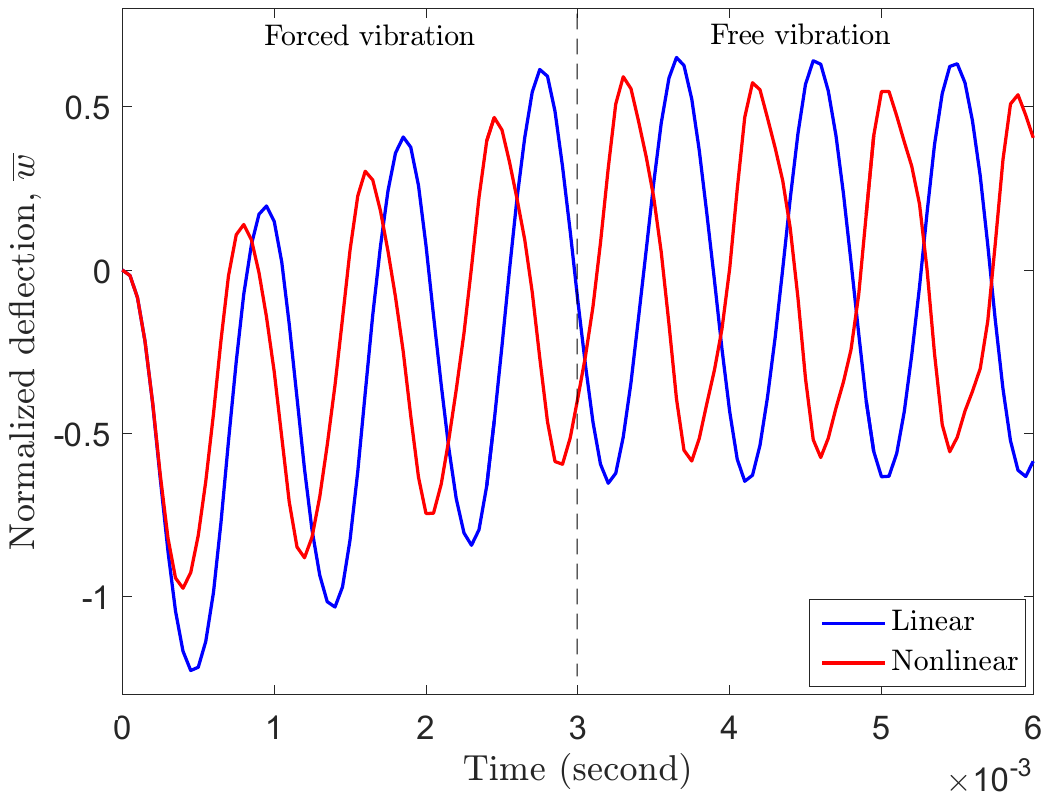}
    \caption{}
  \end{subfigure}
  \hfill
  \begin{subfigure}[c]{0.475\textwidth}
    \centering
    \includegraphics[trim=1.4cm 6.5cm 1.8cm 6.5cm,clip=true,scale=0.47]{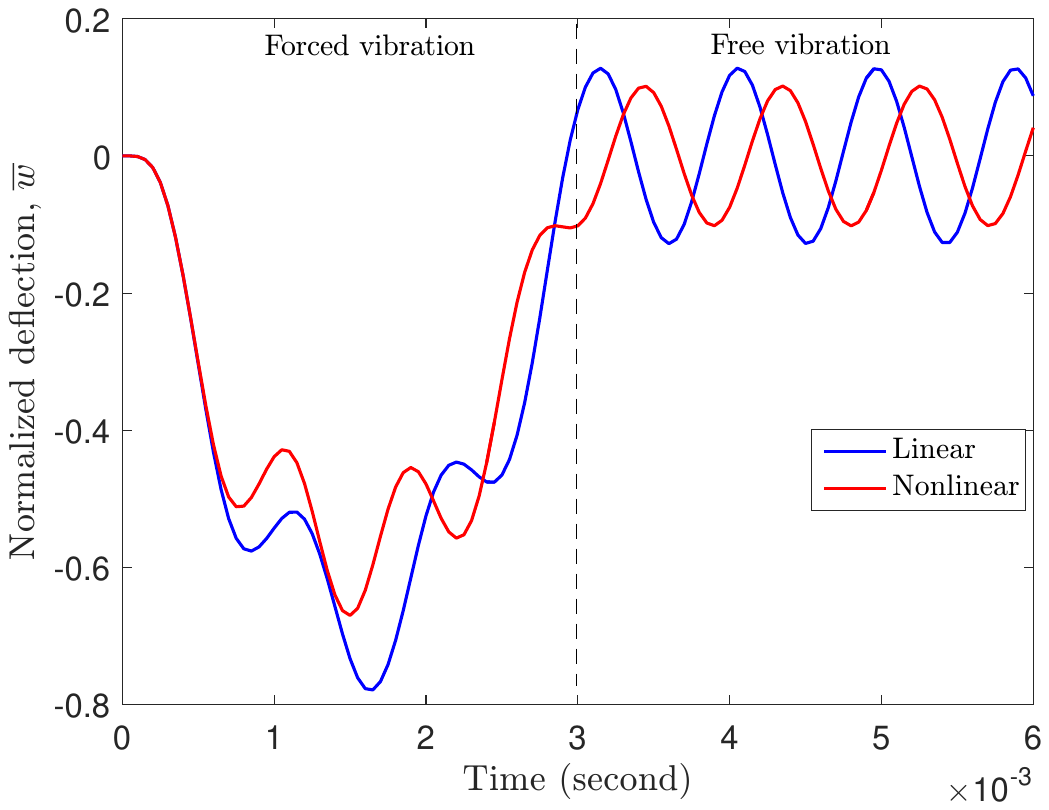}
    \caption{}
  \end{subfigure}
  \hfill
  \caption{Comparison of the linear and nonlinear dynamic responses of CCCC piezoelectric FG porous plate ($a = b = 0.4$ m, $h_c = 20$ mm and $h_p = 1.0$ mm) whose the core layer is constituted by a combination of porosity distribution 2 $(e_0=0.3)$ and GPL pattern C $(\Lambda_{GPL}=1.0 \;wt.\;\%)$: (a) triangular load; (b) sinusoidal load. The FG plate is subjected to dynamic loads, as defined in Eq. (\ref{eqn:load}), in the time interval from 0 to $t_1 = 3 \times 10^{-3}$ second and then is free vibration.}
  \label{fig:Compare_LN}
\end{figure}

Next, the combination influences of various porosity distribution types and GPL dispersion patterns on the nonlinear dynamic behaviour is also examined and indicated in Fig. \ref{fig:Transient_P_GPL}. For this specific example, the porous core layer has porosity coefficient $e_0=0.4$ and GPL weight fraction $\Lambda_{GPL}=1.0\; wt.\;\%$. As clearly demonstrated in Fig. \ref{fig:Transient_P_GPL}, the combination of porosity distribution 1 and GPL dispersion pattern $A$ always provides the best reinforcement as evidenced by obtaining the smallest amplitude of the deflection. Moreover, the linear and nonlinear dynamic responses of FG porous plate with porosity distribution 2 $(e_0=0.3)$ and GPL dispersion pattern $C$ ($\Lambda_{GPL}=1.0\; wt.\;\%$) under triangular and sinusoidal loads are also plotted in Fig. \ref{fig:Compare_LN}. As can be observed, the geometrically nonlinear responses generally obtain smaller magnitudes of the deflection and periods of motion. \\

\subsection{Static and dynamic responses active control}

\begin{figure}[h!]
  \centering
  \includegraphics[trim=0cm 6cm 0cm 6.5cm,clip=true,scale=0.5]{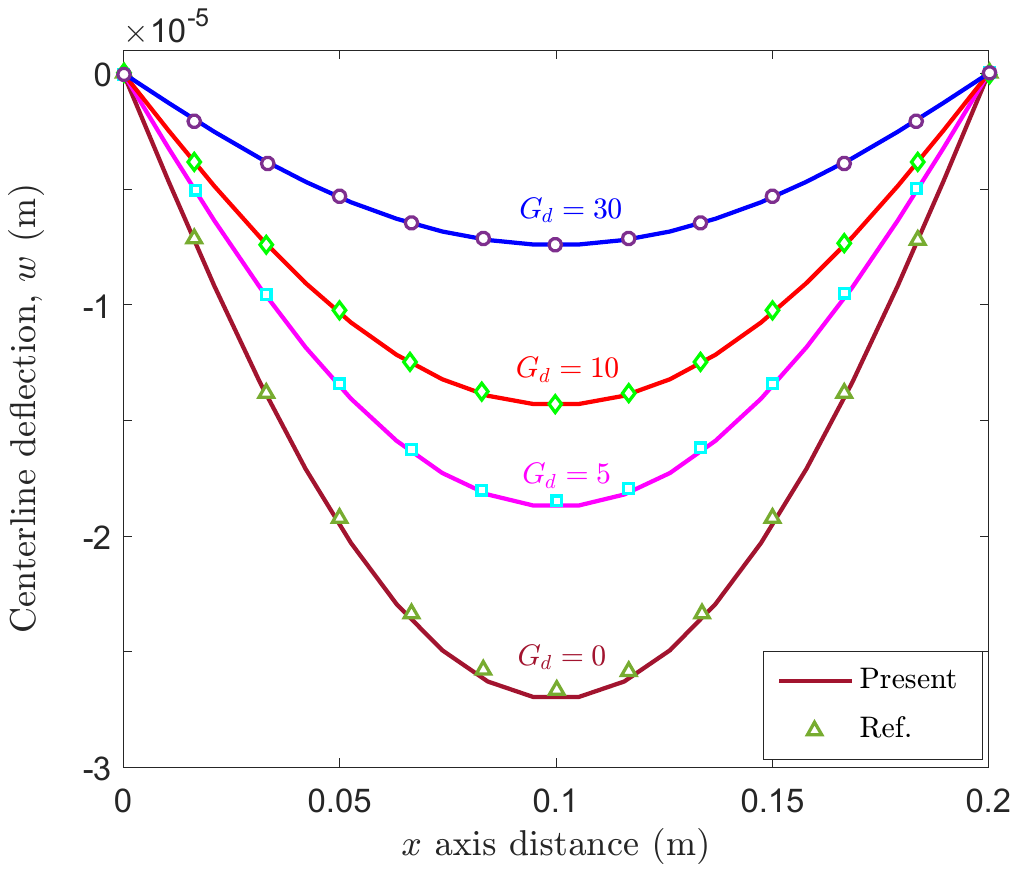}
  \caption{The linear static responses with respect to various values of the displacement feedback control gain $G_d$ for SSSS FGM (Aluminum oxide/Ti-6Al-4V) plate ($a = b = 0.2$ m, $h_c = 1.0$ mm and $h_p = 0.1$ mm, $n = 2.0$) integrated two piezoelectric layers plate under a uniform mechanical load ($q_0 = 100$ N/m$^2$). The value of $G_d = 0.0$ refers to the uncontrolled case and the reference solution is reported in \cite{nguyen2017analysis} using CS-DSG3 based on FSDT.}
  \label{fig:Plot_Gd_compare}
\end{figure}

\begin{figure}[h!]
  \centering
  \includegraphics[trim=0cm 6.5cm 0cm 7cm,clip=true,scale=0.5]{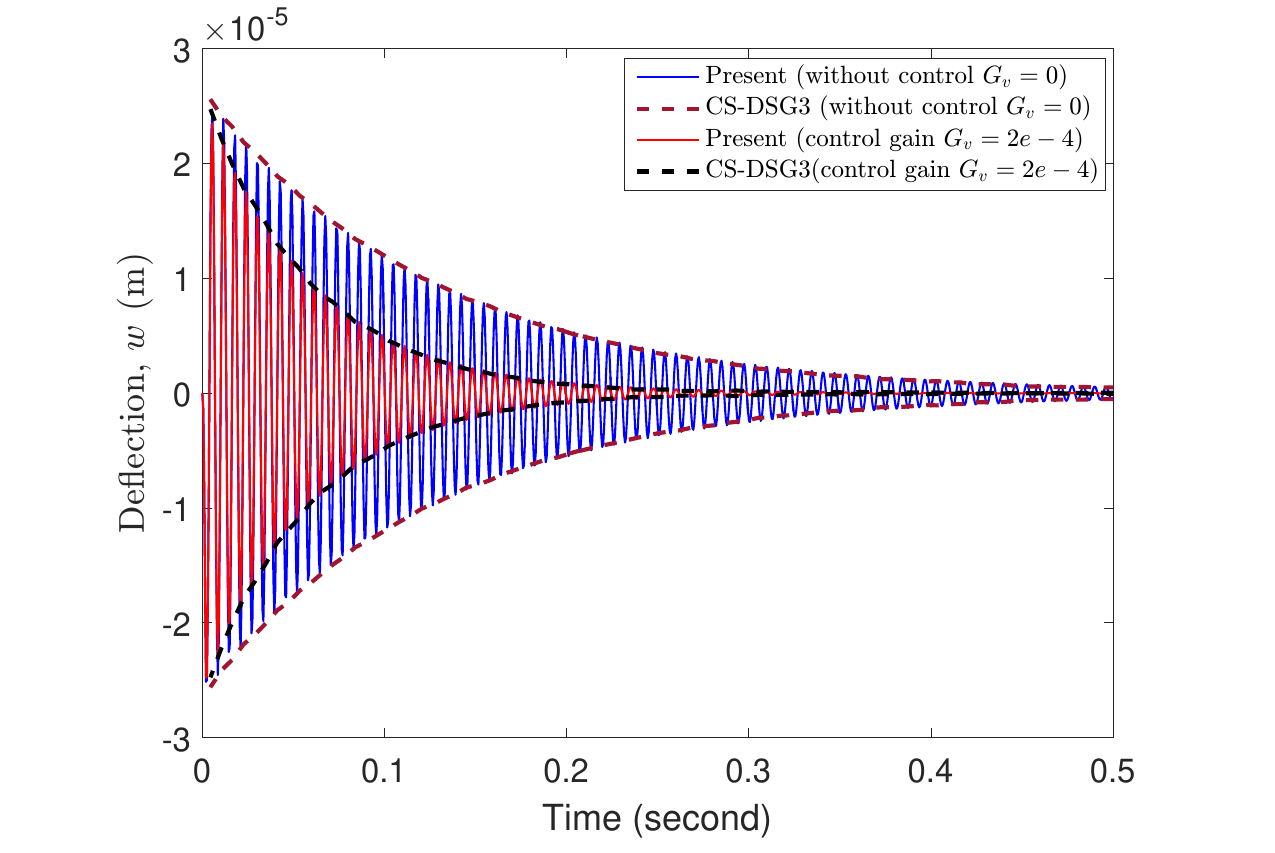}
  \caption{The linear dynamic responses with respect to various values of  the velocity feedback control gain $G_v$ for SSSS FGM (Aluminum oxide/Ti-6Al-4V) plate ($a = b = 0.2$ m, $h_c = 1.0$ mm and $h_p = 0.1$ mm, $n = 2.0$) integrated two piezoelectric layers plate. The FG initially subjected to a uniform mechanical load ($q_0 = 100$ N/m$^2$) and then suddenly eliminated. The value of $G_v = 0.0$ refers to the uncontrolled case and the reference solutions are found in \cite{nguyen2017analysis} using CS-DSG3 based on FSDT. }
  \label{fig:Plot_Lin_Tra_Gd}
\end{figure}

In this section, the active control for static and dynamic behaviors of FG porous plate with GPL reinforcement via sensor and actuator layers is performed. First, to demonstrate the accuracy of the present approach, the active control for linear static behaviors of a SSSS FG plate subjected to a uniform transverse load with $q_0$=100 N/m$^2$ is examined. The FG plate having $a=b= 0.2$ m, $h_c= 1$ mm and $h_p = 0.1$ mm are composed of Ti-6Al-4V and aluminum oxide materials with material index $n=2.0$. Fig. \ref{fig:Plot_Gd_compare} illustrates the linear static deflections of FG plate with various displacement feedback control gains $G_d$. As can be observed that the obtained results from the proposed approach coincide well with the reference solution \cite{nguyen2017analysis} employed the CS-DSG3 based on FSDT. As expected, when the displacement feedback control gain $G_d$ increases, the linear static deflection of FG plate decreases. Furthermore, by using a constant gain velocity feedback $G_v$ and a closed loop control, the active control for the linear dynamic responses of FG plate is also carried out in this work. Accordingly, we consider a FG plate initially subjected to a uniform transverse load $q_0$=100 N/m$^2$ and then suddenly eliminated. It should be noted that to save the computational cost, the modal superposition is exploited and the first six modes are only employed in the modal space analysis. Additionally, for each mode the initial modal damping ratio is assumed to be 0.8 $\%$. Fig. \ref{fig:Plot_Lin_Tra_Gd} illustrates the linear dynamic behavior of central deflection for FG plate corresponding to two values of $G_v$. The results which are generated from present method agree well with the reference solution \cite{nguyen2017analysis}.\\

\begin{figure}[h!]
  \centering
  \includegraphics[trim=0cm 6cm 0cm 7cm,clip=true,scale=0.47]{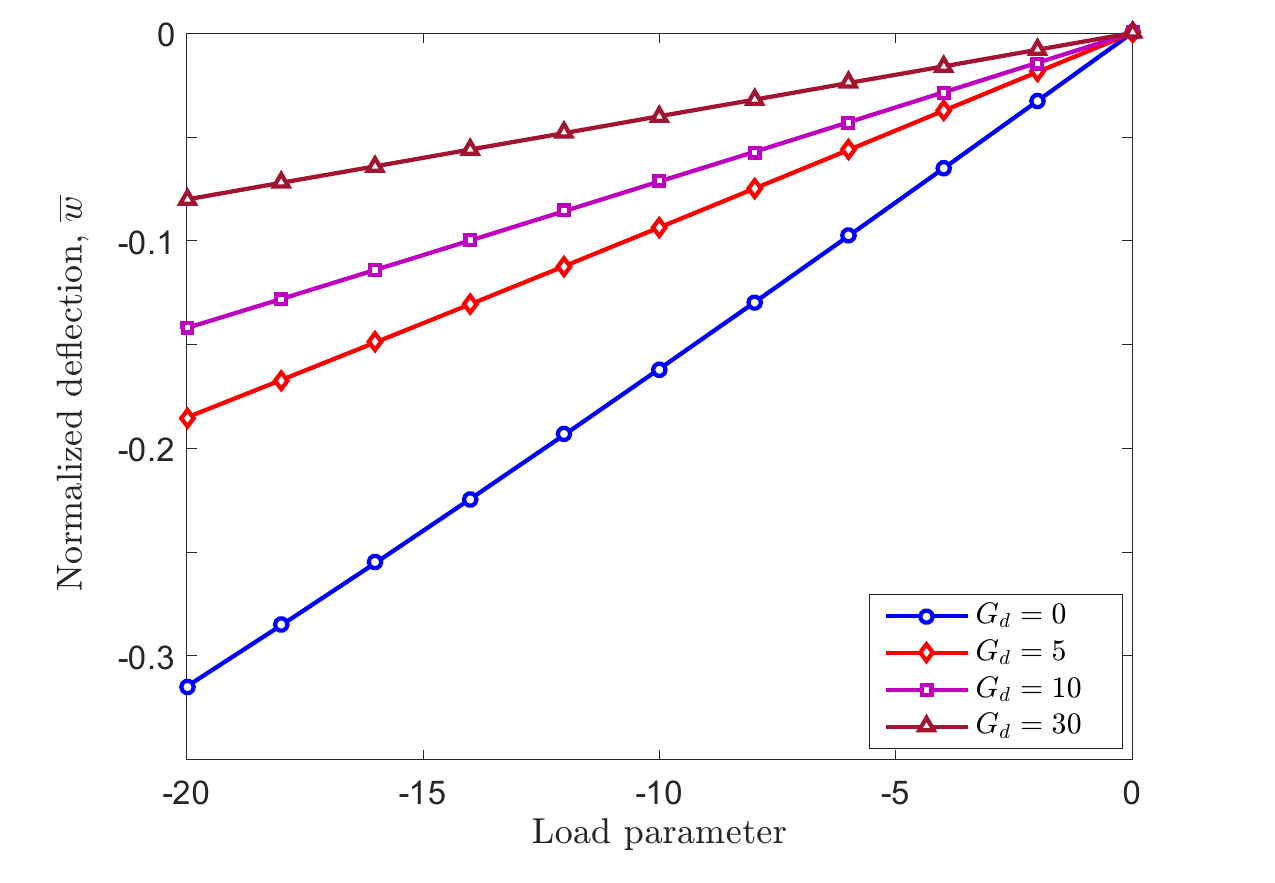}
  \caption{The nonlinear static responses with respect to various values of the displacement feedback control gain $G_d$ for SSSS FG porous plate ($a = b = 0.4$ m, $h_c = 20$ mm and $h_p = 1.0$ mm) with porosity distribution 1 $(e_0=0.4)$ and GPL pattern A $(\Lambda_{GPL}=1.0\;wt.\;\%)$. The value of $G_d = 0.0$ refers to the uncontrolled case. }
  \label{fig:Plot_Non_Gd}
\end{figure}

Next, the active control for nonlinear static responses of a SSSS FG porous plate reinforced with GPLs is further investigated. The core layer of FG plate consisting of combined of porosity distribution 1 and pattern A, which provides the best structural performance, is considered in this example. The material properties are the same in Section \ref{sec:Geo_Static} while the geometric dimensions are given as $a =b= 0.4$ m, $h_c= 20$ mm and  $h_p = 1$ mm. The FG plate is subjected to a sinusoidally distributed load expressed as $q=q_0sin(\pi x/a)sin(\pi y/b)$ with $q_0=1.0$ MPa. Fig. \ref{fig:Plot_Non_Gd} depicts the nonlinear normalized central deflection with $e_0=0.4$ and $\Lambda_{GPL}=1.0\; wt.\;\%$ corresponding to various displacement feedback control gains $G_d$. As can be seen that the normalized central deflection of FG porous plate decreases significantly when the values of displacement feedback control gain increase.\\

\begin{figure}[H]
  \centering
  \begin{subfigure}[c]{0.49\textwidth}
    \centering
    \includegraphics[trim=1.2cm 6.8cm 0.5cm 7cm,clip=true,scale=0.470]{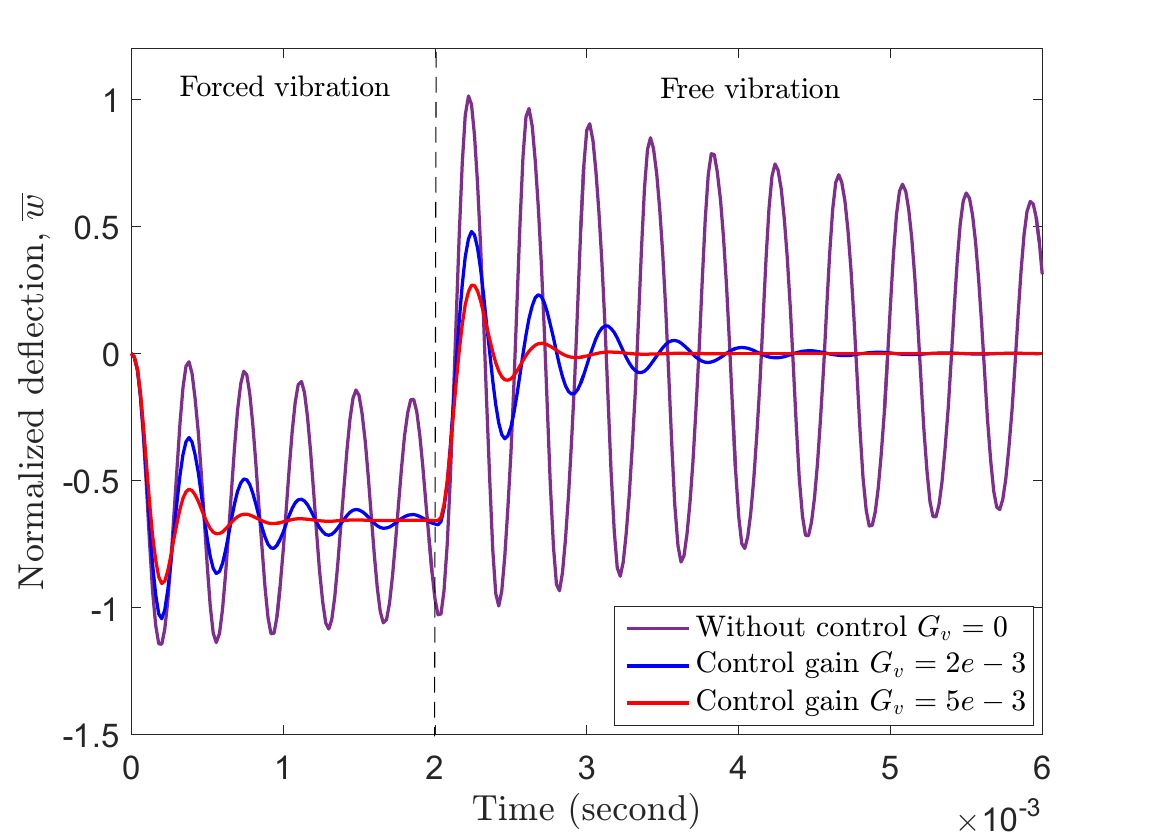}
    \caption{}%
  \end{subfigure}
  \hfill
  \centering
  \begin{subfigure}[c]{0.49\textwidth}
    \centering
    \includegraphics[trim=1.2cm 6.8cm 0.5cm 7cm,clip=true,scale=0.470]{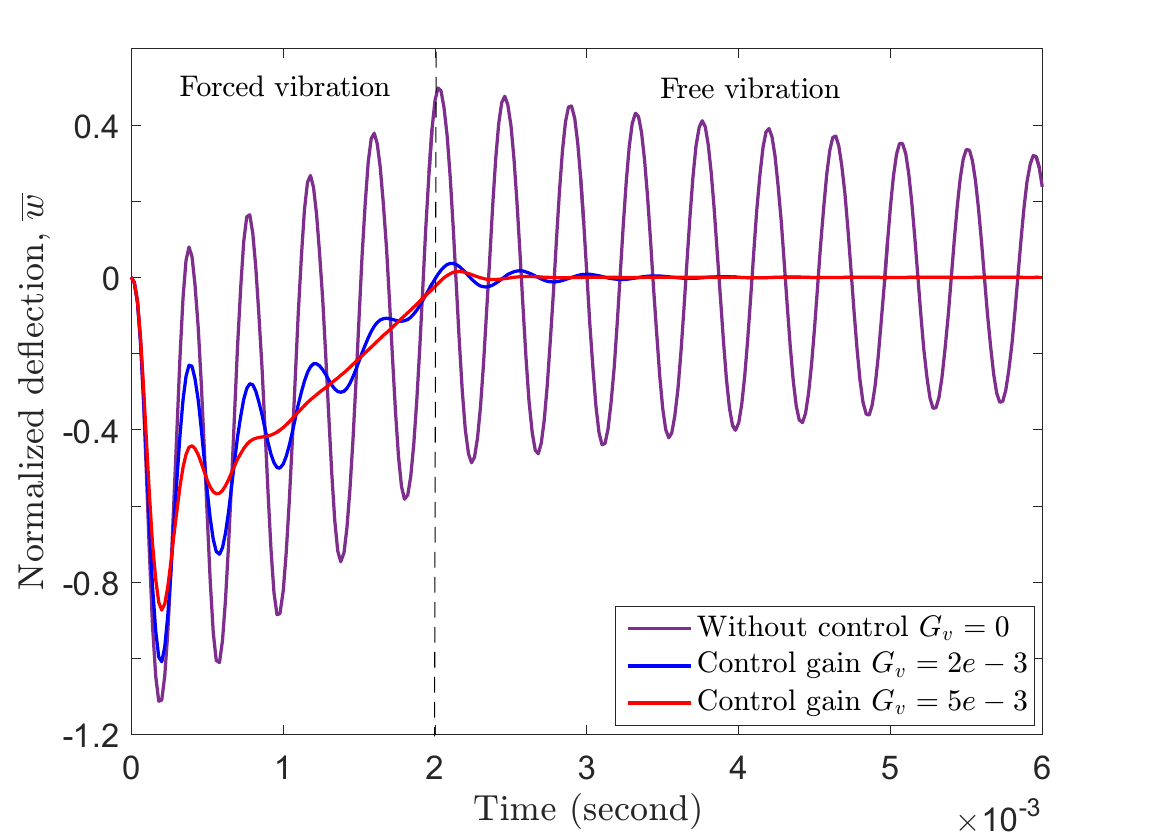}
    \caption{}%
  \end{subfigure}
  \begin{subfigure}[c]{0.49\textwidth}
    \centering
    \includegraphics[trim=1.2cm 6.8cm 0.5cm 7cm,clip=true,scale=0.470]{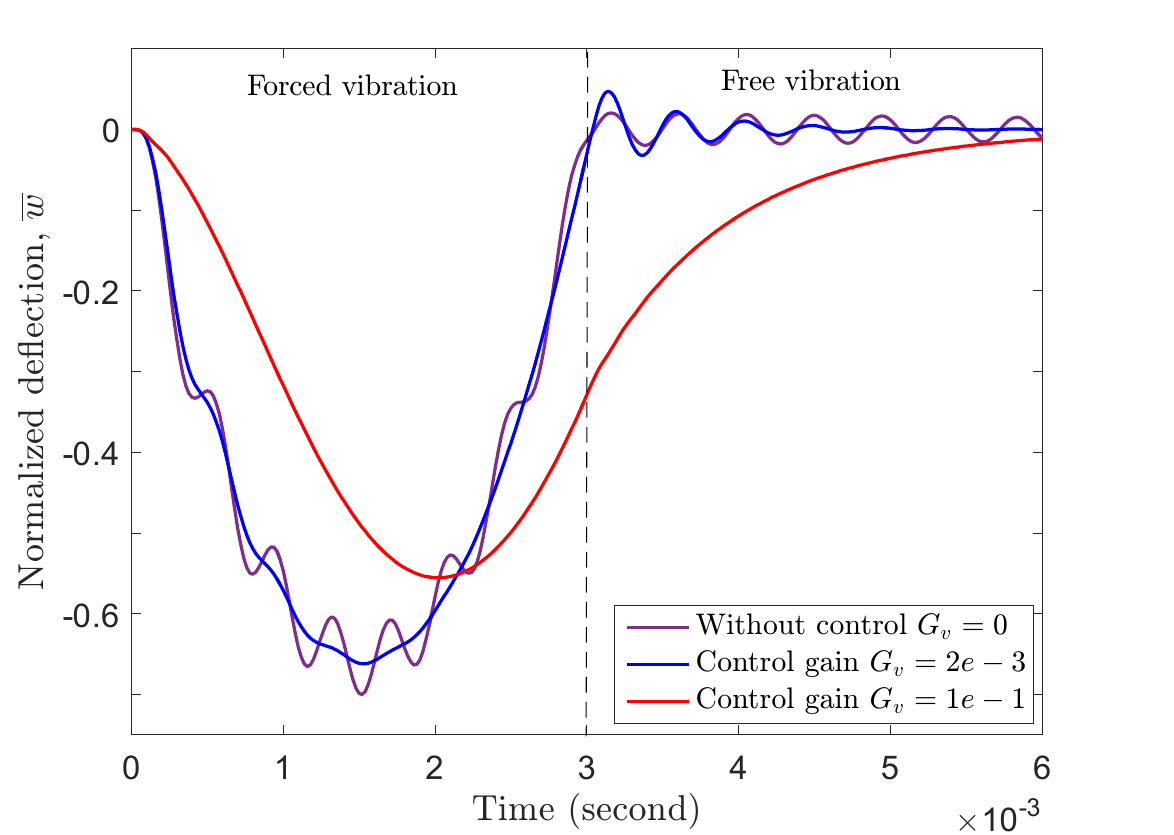}
    \caption{}%
  \end{subfigure}
  \hfill
  \centering
  \begin{subfigure}[c]{0.49\textwidth}
    \centering
    \includegraphics[trim=1.2cm 6.8cm 0.5cm 7cm,clip=true,scale=0.470]{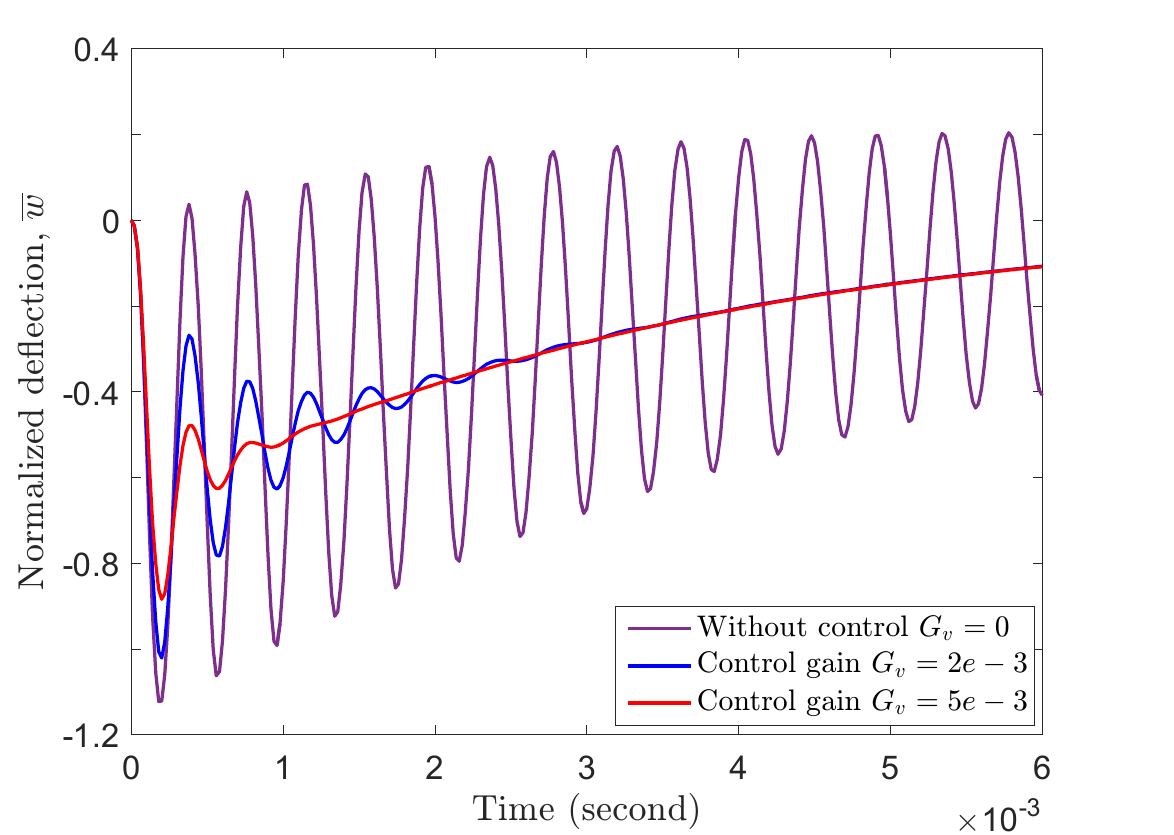}
    \caption{}%
  \end{subfigure}
  \caption{The nonlinear dynamic responses with respect to various values of the velocity feedback control gain $G_v$ for CCCC FG porous plate ($a = b = 0.2$ m, $h_c = 10$ mm and $h_p = 0.1$ mm) with porosity distribution 1 $(e_0=0.4)$ and GPL pattern A $(\Lambda_{GPL}=1.0\;wt.\;\%)$ subjected to dynamic loads: (a) step load; (b) triangular load; (c) sinusoidal load; (d) explosive blast load. The FG plates are subjected to dynamic loads, as defined in Eq. (\ref{eqn:load}), in the time interval from 0 to $t_1$ second and the value of $G_v = 0.0$ refers to the uncontrolled case. }
  \label{fig:Control_Transient_GPL}
\end{figure}

\begin{figure}[H]
  \centering
  \includegraphics[trim=1.2cm 6.8cm 0.5cm 7.5cm,clip=true,scale=0.470]{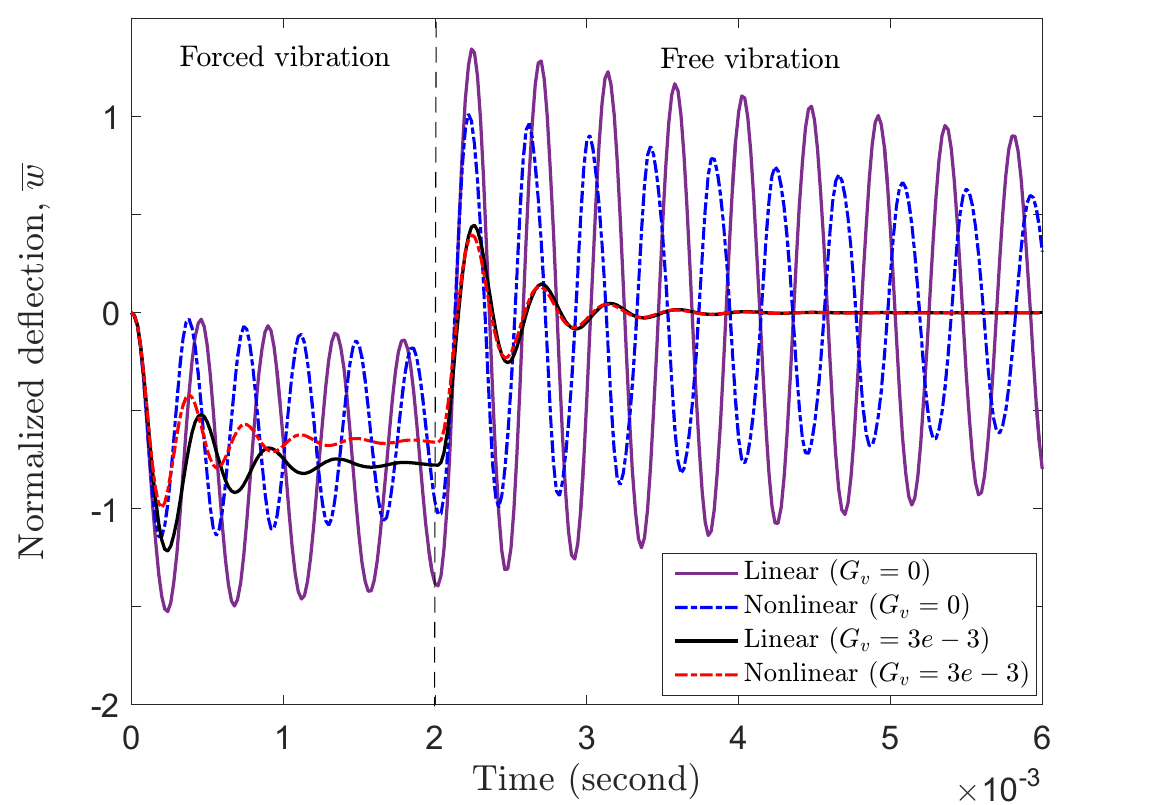}
  \caption{The linear and nonlinear dynamic responses with respect to various values of the velocity feedback control gain $G_v$ for CCCC FG porous plate ($a = b = 0.2$ m, $h_c = 10$ mm and $h_p = 0.1$ mm) with porosity distribution 1 $(e_0=0.4)$ and GPL pattern A $(\Lambda_{GPL}=1.0\;wt.\;\%)$ subjected to step load. The value of $G_v = 0$ refers to the uncontrolled case. The FG plate is subjected to step load, as defined in Eq. (\ref{eqn:load}), in the time interval from 0 to $t_1 = 2 \times 10^{-3}$ second and then is free vibration.}
  \label{fig:Plot_LN_Compare}
\end{figure}

The last example aims to show the active control for geometrically nonlinear dynamic behaviors of a CCCC FG porous plate with GPL reinforcement under sinusoidally distributed transverse loads. The square plate with the porosity distribution 1 $(e_0=0.4)$ and dispersion pattern $A$ $(\Lambda_{GPL}=1.0\; wt.\;\%)$ has $a =b= 0.2$ m, $h_c= 10$ mm and $h_p = 0.1$ mm. Fig. \ref{fig:Control_Transient_GPL} illustrates the nonlinear dynamic responses of normalized central deflection corresponding to various dynamic load types as well as the velocity feedback control gains $G_v$. We observe that in case of uncontrolled ($G_v = 0$), the nonlinear dynamic behaviors of FG porous plate still attenuate with respect to time since the structural damping effect is taken into account in this study. More importantly, the amplitude of central deflection can be suppressed more faster in the case controlled via higher velocity feedback control gain values. As a result, depending on a specific case, the nonlinear dynamic responses of structure such as deflection, oscillation time or even both can be adjusted in order to satisfy an expectation through designing an appropriate value for $G_v$. In addition, Fig. \ref{fig:Plot_LN_Compare} shows the influence of velocity feedback control gain $G_v$ on the linear and nonlinear dynamic responses of plate subjected to step load. As would be expected, the nonlinear dynamic responses obtain smaller magnitudes of central deflection and periods of motion. However, it should be emphasized that the values of feedback control gain could not be enlarged without limit as piezoelectric materials have their own failure voltage values.

\section{Conclusions}
\label{sec:Conclusions}
In this study, the IGA based on B\'ezier extraction of NURBS and $C^0$-HSDT was successfully implemented for the geometrically nonlinear static and dynamic analyses of piezoelectric FG porous plates with GPLs reinforcement. The motion equations were derived based on the $C^0$-HSDT in conjunction with von K\'arm\'am strain assumptions. While the mechanical displacements were estimated by using $C^0$-HSDT based on B\'ezier extraction of NURBS, the electric potential field was assumed as a linear function through the thickness of each piezoelectric layer. Two porosity distributions and three dispersion patterns of GPLs with various related parameters were thoroughly examined via several numerical investigations. The control algorithms based on the constant displacement and velocity feedback were utilized to control the nonlinear static and dynamic responses of FG porous plate structures reinforced with GPLs. A number of key remarks can be drawn from the present work:

\begin{itemize}
  \item  By applying Bernstein polynomials as basis functions in B\'ezier extraction, the IGA can easily be integrated into most existing FEM structures while its advantages are still maintained effectively.

  \item  The stiffness of plate structures is significantly enhanced after adding a small amount of GPLs into metal matrix, while the existence of internal pores yields the reduction in both structural weight and reinforcement effect.

\item The distribution of porosities and GPLs in metal matrix plays a crucial role in reinforcing performance. A combination of porosity distribution 1 with internal pores distributed around the midplane and GPL pattern $A$, where GPLs are close dispersed the top and bottom surfaces, achieve the best reinforcement performance.

  \item The nonlinear dynamic responses can be actively controlled through the velocity feedback control algorithm based on a closed-loop control.

\end{itemize}
\section*{Acknowledgements}
This research was supported by the Grant (NRF-2017R1A4A1015660) from NRF (National Research Foundation of Korea) funded by MEST (Ministry of Education and Science Technology) of Korean government. The third author would like to thank the support of the RISE-project BESTOFRAC (734370)–H2020.

\bibliographystyle{model3-num-names}

\bibliography{Ref_papers}

\end{document}

%% file: preamble.tex
\usepackage{graphicx}
\usepackage{hyperref}
\usepackage{lineno}	
\usepackage{amssymb,bm} 
\usepackage{amsmath} 
\usepackage{navigator}	
\usepackage{changepage}	
\usepackage[pdftex]{color}
\usepackage{caption}
\usepackage{subcaption}
\usepackage{float}
  \usepackage{natbib}

\usepackage{pgfplots}
\pgfplotsset{compat=newest}

\newcommand{\vsigma}{\boldsymbol{\sigma}}
\newcommand{\vvarepsilon}{\boldsymbol{\varepsilon}}

\newcommand{\vkappa}{\boldsymbol{\kappa}}

\newcommand{\vgamma}{\boldsymbol{\gamma}}

\newcommand{\beq}{\begin{linenomath*} \begin{equation}}
\newcommand{\eeq}{\end{equation} \end{linenomath*}}
\newcommand{\bseq}{\begin{linenomath*} \begin{subequations} }
\newcommand{\eseq}{\end{subequations} \end{linenomath*}}

\newcommand{\bal}{\begin{align}}
\newcommand{\eal}{\end{align}}

\newcommand{\bfi}{\begin{figure}}
\newcommand{\efi}{\end{figure}}

\graphicspath{{./figures/}}